\documentstyle[prb,aps]{revtex}
\tighten
\begin{document}
\preprint{IFUM 619/FT  May 7 1998}
\draft
\title
 { Renormalized couplings and scaling correction amplitudes\\
 in the $N$-vector spin models on the sc and the bcc lattices }
\author{P. Butera\cite{pb} and M. Comi\cite{mc}}
\address
{Istituto Nazionale di Fisica Nucleare\\
Dipartimento di Fisica, Universit\`a di Milano\\
16 Via Celoria, 20133 Milano, Italy}
\maketitle
\scriptsize
\begin{abstract}
For the classical $N$-vector model, with  arbitrary $N$,
 we have   computed  through   order
$\beta^{17}$  the high temperature expansions of the second field 
derivative of the susceptibility  $\chi_4(N,\beta)$ on 
the simple cubic  and on the body centered cubic lattices.  
(The  $N$-vector model is also known as the 
$O(N)$ symmetric classical spin Heisenberg  model or, in quantum 
field theory, as the lattice $O(N)$ nonlinear sigma model.)
By analyzing the expansion of $\chi_4(N,\beta)$  on 
 the two lattices, and by carefully
 allowing for the corrections to scaling, 
 we obtain  updated estimates of the critical parameters 
and more accurate tests of the hyperscaling relation 
$d\nu(N) +\gamma(N) -2\Delta_4(N)=0$
 for a range of  values of the spin dimensionality $N$, including
 $N=0$ [the self-avoiding walk model], $N=1$ [the Ising spin 1/2 model], 
 $N=2$ [the XY model],  $N=3$ [the classical Heisenberg model].
Using the recently extended series for 
the  susceptibility and for the second correlation moment,
 we also compute the dimensionless renormalized
 four point  coupling constants and some 
 universal ratios of scaling correction 
amplitudes in fair agreement with recent renormalization group estimates.
\end{abstract}
\normalsize
\narrowtext
 \widetext

\section{Introduction}
We have recently extended the computation of high 
temperature (HT) series for the $N$-vector model\cite{st68} 
with arbitrary spin dimensionality $N$ on the 
$d$-dimensional bipartite lattices, namely 
 on the simple cubic (sc) lattice, 
 on the body centered cubic (bcc) lattice
 and on their $d$-dimensional generalizations.
 In previous papers we 
have tabulated through order $\beta^{21}$ 
the series for the zero field susceptibility $\chi(N,\beta)$ 
and for the second  moment of the correlation function $\mu_2(N,\beta)$ 
 and we have analyzed their critical behavior 
in the $d=2$ case\cite{bc2d} and in the $d=3$ case\cite{bc3d}. 
Here we present a study of the second field derivative 
of the susceptibility  $\chi_4(N,\beta)$  
whose HT expansion on the sc and the bcc
lattices we have extended through order  $\beta^{17}$.
A study of $\chi_4(N,\beta)$ in the $d=2$ case 
had been discussed in Ref.\cite{bc2d}.
 It is interesting to point out  that, in all
analyses presented below, the bcc lattice series appear to be  better 
converged  than the sc lattice series and lead to estimates of
 critical parameters which  are likely  to be more accurate.
 In other words,  the bcc series seem to be   
always "effectively longer"\cite{gut87}
 and therefore give estimates of greater value  than the sc  series.

The list of the  expansions of $\chi_4(N,\beta)$ in $d=3$ 
published up to now is a  short one.
 A decade ago  M. L\"uscher and P. Weisz 
\cite{lw88}(see also Refs.\cite{b90})
 derived  HT expansions of $\chi_4(N,\beta)$ 
through  $\beta^{14}$, for any $N$, 
  on the sc lattice in  $d=2,3,$ and $4$ dimensions by 
  using a linked cluster expansion(LCE) 
technique\cite{lw88,w74mck83,bk,bakeb,rs,re,r1}. 
In the $N=1$ case, [corresponding to the Ising spin 1/2  model]
 the series for the sc lattice published before our work 
already extended through $\beta^{17}$\cite{mackenzie,gutt}
 and has been analyzed by various authors\cite{gutt,ns,roskies}. 
 Finally in the Ising model case,  a series
to order $\beta^{13}$
 on the bcc lattice  and a series 
to order $\beta^{10}$ on the face centered cubic (fcc) lattice
\cite{mackenzie,ns,essam} have long been  available. 

 In our calculation
 we  have  also used  the (vertex renormalized) LCE technique 
and have developed  algorithms which are 
equally efficient in a wide range of space dimensionalities. 
  So far other expansion methods  have 
given competitive (or sometimes superior) performance  only for 
 discrete site variables and  for very simple interactions,
 on  two-dimensional or low coordination number lattices. 
 By the LCE method we have  
 produced  tables of series  expansion 
coefficients  given as explicit functions of the spin dimensionality $N$, 
 with an extension independent of the structure and dimensionality of 
the lattice.
 Thus we have succeeded in  efficiently condensing 
a large body of information  concerning infinitely many universality 
classes. 
 We consider  these coefficient tables  to be 
the main result of our work and, in spite of their considerable extent, 
  we have reported them in the appendix 
 in order to make each step of our work verifiable and  reproducible.
 The size of our computation has been unusually vast: approximately
3$\times 10^6$ topologically inequivalent graphs have  been listed
and evaluated. 
Nevertheless, we are confident that our series 
have been  correctly computed,  
 not only because our codes have  been thoroughly tested, 
but also because 
$N$ and $d$ enter in the whole computational procedure as parameters.
 As a consequence,  at least  simple partial checks 
are available  by observing that  our expansion coefficients,
 when specialized  to $N=1$    
 agree with the  series $O(\beta^{17})$
already available in  3 (as well as  in 2)  dimensions   
and, for $N \rightarrow \infty$,  
 agree with the spherical model\cite{stan2,joy} series 
 which can be readily calculated in  any dimension.
 More comments on the comparison of our results 
with the existing series, can be found in our paper\cite{bc2d} 
 devoted to the two-dimensional $N-$vector model.

 A valuable  justification of our  work is
that an  increasingly accurate study of the critical behavior  
 of  $\chi_4(N,\beta)$ can offer, for all values
of $N$, a sharper test of the 
hyperscaling exponent relation 
$d\nu(N) +\gamma(N) -2\Delta_4(N)=0$.  
Here $\gamma(N)$ and $\nu(N)$ characterize the critical 
singularities in $\chi(N,\beta)$ and $\xi(N,\beta)$ respectively, while 
 $\Delta_4(N)$ is the gap exponent associated with the critical
 behavior of the higher field derivatives of the free energy.
 It is also of great interest to measure accurately 
  the  critical amplitude of $\chi_4(N,\beta)$,   
 which   together with the amplitudes 
of  $\chi(N,\beta)$ and $\xi(N,\beta)$, enters into the 
definition of the universal
dimensionless renormalized four point  coupling constant $g_r(N)$.
Indeed the  uncertainties, probably  still  
 of the order of $1\%$, in the value of this 
 quantity  might   be the main residual 
source of error\cite{mur,gz2}
 in the  present computation of the critical exponents
 within the renormalization group (RG) approach by the 
 Parisi\cite{zib} fixed dimension (FD)
 coupling constant expansion\cite{zib1,brez,zinn,itdr,ant}.
 Murray and Nickel\cite{mur} have recently pushed to seven loop order 
 these calculations and 
 the impact of the 
 additional terms  on the estimates of the critical exponents
 and of some universal amplitude combinations has been critically 
assessed by Guida and Zinn-Justin\cite{gz2}.

As  has been stressed many times in the past two decades
\cite{rogiers,gaunt80,camp80,nickel80,zinn79,adl,fv,fisher,nr90,liu1,liufi}
  and, more recently, also  in Ref.\cite{bc3d}, 
 in order to improve the precision of the   estimates 
 obtainable from HT expansions 
 not only  longer series should be computed, but also
 more careful  allowance should be made for 
 the singular corrections to scaling. 
 Their presence is expected\cite{wegner} and, unsurprisingly,  they
turn out to be important in various cases. 
 Therefore  in this analysis we have 
 also studied their role and have estimated 
   their amplitudes 
in the case of $g_r(N)$,  
both on the sc and the bcc lattice. 
 Moreover, it is of some interest to compute the ratios of these 
 correction amplitudes   with the analogous 
quantities for  $\chi$ and  $\xi$, which  define  interesting 
universal quantities, still subject to significant uncertainty 
 and so far not much studied by HT series methods.
We recall that most
 existing results on the universal 
combinations of the critical and the correction amplitudes are 
reviewed and thoroughly discussed in Refs. \cite{gz2,zinn,aha} 

The paper is organized as follows: In section 
2 we present our notation and 
 define the quantities  we shall study.
 In section 3 we briefly discuss  the simplified 
 doubly biased  differential 
 approximants which  we have used for our estimates beside more traditional
 numerical tools. 
 Our analysis of the series is presented in section 4
along with a comparison to earlier series work, to 
 measures performed in   stochastic simulations and 
to  RG estimates, both by the FD perturbative  technique 
 and by the Fisher-Wilson\cite{epsi}
$\epsilon$-expansion approach\cite{brez,zinn,itdr,epsi1,epv}.
Let us mention that, very recently,  the  $\epsilon$-expansion 
 of $g_r(N)$ has been  extended by Pelissetto and Vicari\cite{pv} from order 
$\epsilon^2$\cite{bgz}  to order $\epsilon^4$, so that we are able to
  compare our HT results also with their estimates. 

 Our conclusions are briefly summarized in section 5.
In the appendix we have reported   the
 HT  series coefficients of $\chi_4(N,\beta)$ 
 expressed in  closed form 
as functions of the spin dimensionality $N$.
For convenience of the reader, we have also reported
 their evaluation for  $N=0$ [the  SAW model\cite{gen}], 
$N=1$  [the Ising spin 1/2 model], 
 $N=2$ [the XY model] and $N=3$ [the classical Heisenberg model].
The present tabulation supersedes and  extends  
the one to order $\beta^{14}$ in Ref.\cite{b90} 
which, unfortunately, contains a few misprints.

\section{ Definitions and notation}

We list here our definitions and notation.
 As the Hamiltonian $H$ of the $N$-vector model we  take:

\begin{equation}
H \{ v \} = -{\frac {1} {2}} \sum_{\langle \vec x,{\vec x}' \rangle } 
v(\vec x) \cdot v({\vec x}').
\label{hamilt} \end{equation}

where $v(\vec x)$ is a  $N$-component classical spin 
of unit length at the lattice site  with position vector $\vec x$,   
and the sum extends to all nearest neighbor pairs of  sites.

The susceptibility is defined by

\begin{equation}
\chi(N,\beta) = \sum_{\vec x} \langle v(0) \cdot v(\vec x) \rangle_c 
 \label{chi} \end{equation}
 
where $\langle v(0) \cdot v(\vec x) \rangle_c$ is 
the connected correlation function between the spin at 
the origin and the spin at the site $\vec x$.

 If we introduce the reduced inverse temperature 
$\tau^{\#}(N) =1 - \beta/\beta^{\#}_c(N) $, 
 (here and in what follows $\#$ stands for either sc or bcc, 
as appropriate,
 then $\chi(N,\beta)$ is expected to behave  like 

\begin{equation}
 \chi^{\#}(N,\beta)
\simeq C^{\#}_{\chi}(N)(\tau^{\#}(N))^{-\gamma(N)}\Big(1+ a^{\#}_{\chi}(N)
(\tau^{\#}(N))^{\theta(N)} +...+ e^{\#}_{\chi}(N)\tau^{\#}(N) +...\Big)
\label{confchi}\end{equation} 

when $ \tau^{\#}(N) \downarrow 0$.
 $C^{\#}_{\chi}(N)$ is the critical amplitude of the susceptibility, 
  $a^{\#}_{\chi}(N)$ is the amplitude of the leading singular correction
 to scaling, $\theta(N)$ is the exponent of this
correction  (also called confluent 
singularity exponent) and 
$e^{\#}_{\chi}(N)$ is the amplitude of the leading regular correction. 
The dots represent higher order singular or analytic correction terms.
The confluent terms result from 
the  irrelevant variables\cite{wegner}.
 Let us recall that not only the critical exponent $\gamma(N)$, but also
 the leading confluent  correction 
exponent $\theta(N)$ is universal (for each $N$). 
On the other hand, the critical 
amplitudes $C^{\#}_{\chi}(N), a^{\#}_{\chi}(N), 
 e^{\#}_{\chi}(N),$ etc. are expected to
 depend  on the parameters of the Hamiltonian and on the lattice structure, 
i.e. they are non-universal.
 Similar considerations also apply  
 to the other  thermodynamic quantities listed below, 
 which have  different critical
 exponents and different critical amplitudes, but
the same leading confluent exponent $\theta(N)$.
 It is known that  $\theta(N) \simeq 0.5$
 for small values of  $N$\cite{zinn}
and, in the context of the large $N$-expansion\cite{ma}, one can 
 also infer that 
 $\theta(N)=  1 +O(1/N)$.

  Since we have clearly stated which quantities 
are universal, from now on we shall 
 generally omit  the superscript $\#$ in order to keep the 
formulas more legible. Notice also that, since there is no chance 
of confusion, we have  systematically omitted the superscript 
$+$ usually adopted for the amplitudes which characterize the high 
temperature side of the critical point. 

The second  moment of the correlation function is defined by

\begin{equation}
 \mu_{2}(N,\beta)=\sum_{\vec x} \vec x^2 \langle v(0) \cdot v(\vec x) 
\rangle_c 
\end{equation}

In the vicinity of the critical point $\mu_{2}$ is expected to 
behave  as 

\begin{equation}
 \mu_{2}(N,\beta)
\simeq C_{\mu}(N)\tau^{-\gamma(N)-2\nu(N)}\Big(1+ a_{\mu}(N)
\tau^{\theta(N)} +...
+ e_{\mu}(N)\tau +...\Big)
\label{confmu}\end{equation} 
 as $ \tau \downarrow 0$.

In terms of $\chi$ and $\mu_{2}$,  the 
 second moment correlation length $\xi$ is defined\cite{tarko} by 

\begin{equation}
 \xi^{2}(N,\beta)= \frac  {\mu_{2}(N,\beta)} {6\chi(N,\beta) }. 
\end{equation}

In the vicinity of the critical point $\xi$ 
is expected to behave  as 

\begin{equation}
 \xi(N,\beta)
\simeq C_{\xi}(N)\tau^{-\nu(N)}\Big(1+ a_{\xi}(N)
\tau^{\theta(N)} +...
+ e_{\xi}(N)\tau +...\Big)
\label{confxi}\end{equation}
 as $ \tau \downarrow 0$.

The second field derivative of the
susceptibility is defined by
\begin{equation}
 \chi_{4}(N,\beta )={\frac {3N} {N+2}}  \sum_{x,y,z}
\langle  v(0) \cdot v(x) v(y) \cdot v(z)\rangle_{c}=
{\frac {3N} {N+2}} \Big(-\frac {2} {N} 
+ \sum_{r=1}^\infty d_r(N) \beta^r \Big).
\label{chi4}\end{equation}
 Notice that this definition differs by a factor $1/N^2$ from 
that used in Ref.\cite{lw88}.

It is well known\cite{stan2,joy} that, for $N \rightarrow \infty$ 
at fixed $\tilde \beta \equiv \beta/N$, $\chi(N, \beta )$ has a finite
non trivial limit $\tilde\chi(\tilde \beta)$. 
 On the other hand, as expected,  in the same limit
we have $\chi_{4}(N, \beta )= O(1/N)$.
 It is  the quantity $N\chi_{4}(N, \beta)$ 
 that has a finite limit 
$\tilde \chi_4(\tilde \beta)$ simply  expressed as
\begin{equation}
\tilde \chi_4(\tilde \beta)=
-6\tilde\chi^2(\tilde \beta)\big(\tilde\chi(\tilde \beta)
+\tilde\beta \frac {d\tilde\chi(\tilde \beta)} {d\tilde\beta}\big).
\label{chi4sfer}\end{equation}
 
  Also the  
$N \rightarrow 0$  limit, at fixed  $\tilde \beta$, exists\cite{gen} 
and the quantity
\begin{equation} 
\hat \chi_{4}(\tilde \beta) = \lim_{N \to 0} \chi_{4}(N, \beta ) =
 -3 \sum_{N_1,N_2} c_{N_1 N_2} \tilde \beta^{N_1+N_2}   
\label{chi4saw}\end{equation}
 has the following 
interpretation\cite{fernandez,lms}: $c_{N_1 N_2}$ is the number of pairs 
 $(\omega^{(1)}, \omega^{(2)})$ of self avoiding walks
such that 
$\omega^{(1)}$ is a $N_1$-step walk starting at the origin and 
$\omega^{(2)}$ is a $N_2$-step walk starting anywhere and crossing 
$\omega^{(1)}$. 

In the vicinity of the critical point 
$\chi_{4}(N,\beta)$ is expected to behave  as 

\begin{equation}
 \chi_{4}(N,\beta)
\simeq C_{4}(N)\tau^{-\gamma(N)-2\Delta_4(N)}\Big(1+ a_{4}(N)
\tau^{\theta(N)} +...
+ e_{4}(N)\tau +...\Big)
\label{conf}\end{equation}
 as $ \tau \downarrow 0$.

In terms of $\chi$, $\xi$  
and $ \chi_{4}$  the
"dimensionless renormalized
four point coupling  constant" $ g_r(N)$ is defined 
 as the value  of

\begin{equation}
 g(N,\beta)\equiv - \frac{v f(N) \chi_{4}(N,\beta)}
{\xi^3(N,\beta) \chi^2(N,\beta)}
\end{equation}

at the critical point $\beta_c(N)$.
Here $f(N)=\frac {N+8} {48\pi} $ is a normalization factor 
chosen in order to match the usual field theoretic definition 
of $g_r(N)$\cite{zinn} and  $v$ denotes the volume per lattice 
site ($v=1$ for the sc 
lattice and $v=4/3\sqrt3$ for the bcc lattice).

In the vicinity of the critical point $g(N,\beta)$ is 
expected to behave  as 

\begin{equation}
  g(N,\beta)
\simeq  g_r(N)\tau^{\gamma(N)+3\nu(N)-2\Delta_4(N)}\Big(1+ a_{g}(N)
\tau^{\theta(N)}+...
+ e_{g}(N)\tau +...\Big)
\label{confg}\end{equation} 
 as $ \tau \downarrow 0$, with

\begin{equation}
  g_r(N)
=  - \frac{v f(N) C_{4}(N)}{C^3_{\xi}(N) C^2_{\chi}(N)}
\label{ampg}\end{equation} 

The  Gunton-Buckingam\cite{gunt,rig,glja} inequality 
\begin{equation}
3\nu(N) + \gamma(N) -2\Delta_4(N) \geq 0 
\label{gunt}\end{equation}

 together with the Lebowitz\cite{lebo} inequality 
$\chi_4(N,\beta) \leq 0$, 
implies that $ g(N,\beta)$ is a bounded non-negative 
 quantity as $\tau \downarrow 0$. 
 The  vanishing of $ g(N,\beta_c)$   is a sufficient condition
for Gaussian behavior at criticality, or, in lattice field theory
language,
for "triviality" \cite{fernandez} of the continuum field 
theory defined by the  $N-$vector lattice model in the critical
 limit. If  $\chi_4(N,\beta)$ is nonvanishing  and 
 the above inequality holds as an equality (the hyperscaling relation)

\begin{equation}
3\nu(N)+ \gamma(N)  -2\Delta_4(N) = 0
\label{hyp}\end{equation} 
then 
\begin{equation}
  g(N,\beta)
\simeq  g_r(N)\Big(1+ a_{g}(N)
\tau^{\theta(N)}+...
+ e_{g}(N)\tau +...\Big)
\label{confgg}\end{equation} 

namely $ g(N,\beta)$  tends to the nonzero 
limiting value  $g_r(N)$ as $\tau \downarrow 0$.

For checking purposes it is useful 
 to recall here   the large $N$ limits of
 the critical  amplitudes.
They have been computed\cite{joy}  long
 ago 
\begin{equation}
C^{sc}_{\chi}(\infty)
=\frac {1} {16\pi^2(\tilde \beta^{sc}_c(\infty))^3}=0.39228768..
\label{ampsfersc}\end{equation}
 with $\tilde \beta^{sc}_c(\infty)= 0.2527310098..$ and
\begin{equation}
C^{bcc}_{\chi}(\infty)
=\frac {1} {64\pi^2(\tilde \beta^{bcc}_c(\infty))^3} =0.29974101..
\label{ampsferbcc}\end{equation}
with $\tilde \beta^{bcc}_c(\infty)= 0.1741504912..$ 

Moreover, we recall that, since in the large $N$ limit
\begin{equation}
\tilde \mu_2= q\tilde \beta \tilde \chi^2
\label{xichisfer}\end{equation}
 where $q$ is the lattice coordination number,  we have 
$C^{\#}_{\xi}(\infty)
=(q\tilde\beta^{\#}_c(\infty) C^{\#}_{\chi}(\infty)/6)^{1/2}$. 
 On the other hand, if we  denote by $\tilde C^{\#}_{4}(\infty)$ 
 the large $N$ limit of  $NC^{\#}_{4}(N)$, by (\ref{chi4sfer})  we have 
$\tilde C^{\#}_{4}(\infty)=-12(C^{\#}_{\chi}(\infty))^3$ and therefore it 
 follows that $g_r^{\#}(\infty)=1$.

\section{ Analysis of the series} 
 As  mentioned in the introduction,   
a variety of careful analyses
\cite{rogiers,gaunt80,camp80,nickel80,zinn79,adl,fv,fisher,nr90,liu1,liufi} 
of the Ising model HT expansions  as well as 
 our  study of the recently extended  $N$-vector model series\cite{bc3d},
 suggest that the   non-analytic confluent 
corrections to the leading critical behavior of 
 the thermodynamic quantities, indicated 
in the asymptotic formulas(\ref{confchi}),(\ref{confmu}), 
(\ref{confxi}), etc.  exist and should not in general be 
neglected in computing numerical estimates of the critical parameters.
 It has long been observed
\cite{rogiers,gaunt80,camp80,nickel80,zinn79,adl,fv,fisher,nr90,liu1,fereconf} 
that these corrections reveal themselves as small 
apparent violations of both 
universality and hyperscaling in a naive pure power law analysis
 of the critical behavior.
However it is  also well 
known \cite{nickel80,zinn79}
 that, unless very long HT series are available,   extracting 
simultaneously estimates for
 $\beta_c$, the exponents and the amplitudes of  the critical {\it and}
 of the subleading singularity 
 is a  difficult and unstable 
 numerical problem.
For this task the inhomogeneous  
DA  method\cite{gutasy} 
of series analysis is generally believed to be  more effective than 
 the traditional and simpler
Pad\'e approximant (PA) method, because, at least in principle,
 it might be flexible 
enough to represent functions behaving 
like $\phi_1(x) (x-x_0)^{-\omega} +
\phi_2(x)$ near a singular point $x_0$, where $\phi_1(x)$
is a   regular function of $x$  and $\phi_2(x)$ may 
contain a (confluent)
singularity of
strength smaller than $\omega$.
 Unfortunately, in practice,  this is not completely true:
 very long series are  needed  anyway
and/or the procedure should be biased  by choosing
 very carefully  the structure of the 
approximants     and
 by giving proper inputs.
 We have followed here the latter approach. 
 As in some of our previous studies\cite{bc3d,bcpre},
 beside more standard procedures of analysis, 
  we have  employed   a  doubly biased prescription
 which assumes 
that the confluent exponent $\theta$  and the 
inverse critical temperature $\beta_c$ are accurately known. 
 This procedure seems to perform reasonably 
 well, even with not very long series. 
We  have taken the values of $\theta(N)$ as estimated by
the FD renormalization group method.  More precisely, for  $N \leq 4$, 
 we have used the values  suggested
by Guida and Zinn-Justin\cite{gz2}, and for 
 $N>4$, we have used the  six loop
estimates   recently obtained  
 by A. I. Sokolov\cite{sokoth} and kindly communicated to us before 
publication. These values  are reported in 
table 1.
We also have assumed that the critical temperatures $\beta_c^{\#}(N)$
 have  been  determined accurately enough in our previous 
study of the susceptibility\cite{bc3d}.

Let us now  recall in some detail
 the  features of the 
 simplified  DA  method.

 We wish to approximate some function, given as a series expansion
 around $\beta=0$   and expected,  
when $\beta \uparrow \beta_c$,
to have the form
\begin{equation} 
 f(\beta)= \sum_{n=0} f_n \beta^n \simeq b(\beta) 
+ c(\beta) (1-\beta/\beta_c)^{\theta} 
+ o\big( ( 1-\beta/\beta_c)^{\theta})\big). 
\label{fbe}\end{equation} 

We assume that $ \beta_c $ and the  real positive exponent $\theta$  
are accurately known, and that  
 $b(\beta)$ and $c(\beta)$  are analytic at $\beta=\beta_c$.  We set 
$b(\beta_c)=b_0$ and $c(\beta_c)=c_0$.
 
We shall  estimate the function $f(\beta)$ and therefore the constants
$b_0$ and $c_0$
 by the particular class of first order inhomogeneous  differential 
 approximants  
 $F(\beta)$ defined as the solutions of the equations
\begin{equation} 
Q_m(\beta)\Big[(1-\beta/\beta_c)\frac {dF(\beta)}{d\beta}+
\frac{ \theta} {\beta_c} F(\beta)\Big]
+R_n(\beta)=0
\label{damod}\end{equation} 
with the initial condition $F(0)=f_0$.

 $Q_m(\beta)$ and $R_n(\beta)$ are  polynomials of degrees $m$ and $n$
 respectively, whose coefficients are calculated, as usual, 
by imposing that the  series 
expansion of $F(\beta)$ agrees with that of 
$f(\beta)$ at least through  the order 
$\beta^{m+n+1}$. In addition the normalization condition $Q_{m}(0)=1$ is 
imposed.
 Assuming for simplicity $0 < \theta < 1$, 
  $f(\beta_c)=b_0$ is  estimated as 

\begin{equation} 
b_0^{(n,m)} = \frac {-\beta_c R_n(\beta_c)} {\theta Q_m(\beta_c)}
\label{b0nm}\end{equation} 

and the amplitude $c_0$ of the subleading term in Eq. (\ref{fbe}) 
 is estimated    by the formula
\begin{equation} 
 c_0^{(n,m)}=f_0-b_0^{(n,m)}
   - \int_0^{\beta_c}\frac { D^{(n,m)}(t)dt}
{ {(1 - t/\beta_c)}^{1+\theta} } 
 \label{amps} \end{equation} 

where
\begin{equation}
D^{(n,m)}(t)= \frac{R_n(t)} {Q_m(t)} - 
\frac {R_n(\beta_c)} { Q_m(\beta_c)}
 \label{ampsd} \end{equation}

We shall  consider only  the "almost diagonal" approximants with
 $\vert m-n \vert \le 4$.

 The approximants defined by (\ref{damod})
 are just a small subclass of the 
general first order inhomogeneous DA's

\begin{equation} 
 (1 - \beta/\beta_c )Q_m(\beta)\frac {dF(\beta)}{d\beta}
 +P_l(\beta)F + R_n(\beta)=0
 \label{dahb1} \end{equation} 

biased with $\beta_c$ 
and with $\theta$ 
 by imposing $P_l(\beta_c)/ Q_{m}(\beta_c)=\frac{\theta} {\beta_c}$.
 Still assuming $0<\theta <1$, we can estimate
 $b_0$ and $c_0$  from 
(\ref{dahb1}) as follows

\begin{equation}
 b_0^{(m;n,l)}= -\frac {R_n(\beta_c)} {P_l(\beta_c)}
=-\frac{\beta_c R_n(\beta_c)} {\theta Q_m(\beta_c)}
\label{dab0} \end{equation}

\begin{eqnarray}
 c_0^{(m;n,l)}=-b_0^{(m;n,l)} + g^{(m;n,l)}
(\beta_c)\Big[ f_0
-\int_0^{\beta_c}\frac {D^{(m;n,l)}(t)}
 {g^{(m;n,l)}(t)(1-t/\beta_c)^{1+\theta}}
\nonumber \\
 + \frac{\theta} {\beta_c}
b_0^{(m;n,l)}\int_0^{\beta_c}\big(\frac{1} {g^{(m;n,l)}(t)}-\frac{1} 
{g^{(m;n,l)}(\beta_c)}\big) \frac{dt} {(1-t/\beta_c)^{1+\theta}}\Big]
\label{dac0} \end{eqnarray}

where

\begin{equation}
 g^{(m;n,l)}(\beta)= exp\Big[ - \int_0^{\beta}\big( \frac{P_l(t)} {Q_m(t)}-
\frac{P_l(\beta_c)} {Q_m(\beta_c)}\big) \frac{dt} {(1-t/\beta_c)}\Big]
\label{gab0} \end{equation}

 and $D^{(m;n,l)}(t)$ has the same form as (\ref{ampsd}).
 The simple formulas  (\ref{b0nm}) and  (\ref{amps}) 
 are  recovered from the general formulas (\ref{dab0}) and  (\ref{dac0})
by subjecting $P_l(\beta)$ to the further strong constraint 
$P_l(\beta) \equiv \frac{ \theta}{\beta_c} Q_{m}(\beta)$. 
This prescription, which, for short, we will refer to as simplified 
differential approximants  (SDA's) might also be viewed as a  simple DA-like
 generalization of the biased PA 
method introduced in Refs.\cite{rosk,adp,nd}.

We have carried out many numerical experiments on simple model series 
 having the analytic structure (\ref{fbe}).
 They show that the SDA's, when biased with the exact values of 
$\beta_c$ and $\theta$, are able to  produce very accurate estimates 
 of $b_0 $ and fairly accurate estimates 
of the confluent amplitude $c_0$.
 In practice however, we do not have strict control on the series:  
 only approximate values of $\beta_c$ and $\theta$
are available for biasing the SDA's and we do not 
 know the strength of the subleading correction terms and of the smooth 
 background. Therefore it is important 
  to understand how sensitive 
 are the estimates of $b_0 $ and $c_0$ to the errors 
 in the biased inputs and how they depend on the structure 
of the singularity.
It turns out that the estimates of $b_0$ 
are rather stable  
 when the biased value   
 for $\beta_c$  and for $\theta$ are varied away from their 
 true values
in a range comparable to the 
typical estimated uncertainties in the realistic cases. 
On the other hand,  $c_0$ 
appears  to be  much more sensitive 
to errors in the biased values. Let us consider, 
 to be definite, the case of  the very simple test series
\begin{equation} 
 f(\beta)=  c_0(1 - \frac{\beta} {\beta_c} )^{\theta} 
+ c_1(1 -\frac{\beta} 
{\beta_c })^{2\theta} + b_0 exp(1-\frac{\beta} {\beta_c}) 
 \simeq
 b_0 + c_0(1-\frac{\beta} {\beta_c} )^{\theta} 
 + o((\beta_c-\beta)^{\theta})
 \label{testser} \end{equation} 
 which we have examined for various values of  $\theta$.
 If  the size of the subleading correction to scaling is much smaller than
 the size of the leading one, namely if 
$\mid c_1\mid \ll\mid c_0\mid$ and 
 we bias the calculation 
with the exact values of the parameters $\theta$ and $\beta_c$, 
 we  are able to determine $b_0$  by (\ref{b0nm}) to within less 
than $10^{-2}\%$ and  
$c_0$ by (\ref{amps})  to within less than $1 \%$.   
 However, if the SDA's  
are biased with a value of $\theta$  which 
is off the right value by $5\%$, then the relative 
error of $c_0$ can  become  as large as $15\%$,  
while the error of $b_0$
 increases to some $0.1\%$.
The precision of $b_0$ remains essentially unchanged, but the
sensitivity of $c_0$ to variations in the biased
 values and, as a consequence,  the accuracy of
 its estimate is somewhat worsened 
 in the slightly more complicated, but sometimes
 realistic case in which
 $\mid c_1\mid \approx \mid c_0 \mid$.
 Unsurprisingly, the worst situation occurs when   the leading confluent 
amplitude is much smaller than the subleading one, 
since   the uncertainty in the numerical estimate of 
 $c_0$  may then become very large.
In conclusion, taking a  conservative
attitude, we can safely expect that, for the HT series we are 
going to study, the relative error 
 on the value of $f(\beta)$ at $\beta_c$  can be much smaller than $1\%$, 
 while the uncertainty of the
correction amplitude can be as large as 
 $20\%$,  unless the amplitude 
 is  very small: in this case, due to  
 a higher  sensitivity 
to the biased values and/or to the neglect of possibly  
 important  subleading 
corrections, our estimates are  likely to be much more inaccurate.
In order to better understand these results let us also
 observe that, if we tried to estimate $b_0$ in (\ref{testser})
 by simple PA's biased with
$ \beta_c $, the relative error would be substantially 
larger and increasing with the size of the correction amplitude.
 Finally, we remark that in all computations  presented below, 
the error estimates are always somewhat subjective. They include effects
 both from  the scatter of the approximant values,  possible 
 residual trends in sequences of estimates,
 as well  uncertainties of the bias inputs.  

 We have applied  the SDA approximation procedure  not only to the 
 quantity $g(N,\beta)$ in order to compute 
the confluent amplitude $a_g(N)$, but also to the "effective exponent"
 of $\chi_4$

\begin{equation} 
 \gamma_4(N,\beta) \equiv (\beta_c(N)-\beta) 
\frac {d{\rm log}\chi_4(N,\beta)} {d\beta} =
 \gamma(N)+2\Delta_4(N) - a_{4}(N)\theta(N)\tau^{\theta(N)} 
+ o\big( \tau^{\theta(N)}\big) 
\label{gamma4}\end{equation} 
 in order to compute the critical exponent and the 
confluent amplitude $a_{4}(N)$.

 Moreover we have examined the analogous quantities

\begin{equation} 
 \gamma(N,\beta) \equiv (\beta_c(N)-\beta) 
\frac {d{\rm log}\chi(N,\beta)} {d\beta} =
 \gamma(N) - a_{\chi}(N)\theta(N)\tau^{\theta(N)} 
+ o\big( \tau^{\theta(N)}\big) 
\label{gamma}\end{equation} 

 in order to compute the confluent amplitude $a_{\chi}(N)$,
 and 
\begin{equation} 
\nu(\beta,N) \equiv  {1\over 2} (\beta_c(N)-\beta) 
\frac {d{\rm log}\big [ \xi^2(N,\beta)/\beta\big ]} {d\beta}=
\nu(N) - a_{\xi}(N)\theta(N)\tau^{\theta(N)}
+ o\big( \tau^{\theta(N)}\big)
\label{nu}\end{equation} 

 in order to compute $a_{\xi}(N)$.
 Notice that the
estimates thus obtained for the confluent amplitudes 
 $a_{\chi}$,  $a_{\xi}$, and $a_4$
  are biased solely with 
$\beta_c$ and $\theta$. 
However, due to their definition   as residua, the sensitivity
 of the results to the biased value for $\beta_c $ is higher than in the 
case of $g_r$.

 The estimates of the 
critical amplitudes have been 
 obtained 
 by examining  quantities like 
\begin{equation} 
\tau^{\gamma(N)} \chi(N,\beta)
\simeq C_{\chi}(N)\Big(1+ a_{\chi}(N)
\tau^{\theta(N)} +...+ e_{\chi}(N)\tau +...\Big)
\label{nubb}\end{equation} 
 or the analogous expressions for $\chi_4$ and $\xi^2$. 
 This procedure  also yields the correction amplitudes, but 
since it  
  requires biasing also with
 the critical exponent $\gamma(N)$ (or $\nu(N)$ etc.),
 we expect that the corresponding 
 results will be subject to  a larger uncertainty.

In  conclusion, whenever  sizable confluent corrections  
are present, the doubly  biased SDA procedure will 
produce values of $g_r(N)$ which are slightly, but definitely different 
 from  estimates by generic DA's not 
directly constrained to reproduce the 
confluent singularity and, {\it a fortiori},  from the simple PA estimates.
Indeed, since $\theta < 1$, the function $g(N,\beta)$ will approach 
with a divergent slope
 its value at $\beta_c(N)$, from above
    if the correction amplitude is positive
 or,  otherwise, 
 from below.
 As a consequence,   too smooth 
extrapolations of $g(N,\beta)$  to the critical point 
 $\beta_c$ 
 would overestimate the correct result in the 
former case and underestimate it in the latter.
Analogous problems will occur in the study 
of the exponents and of the correction 
amplitudes for $\chi,\mu_2,\chi_4$, the only difference being that, since
in the formulas for the effective exponents 
(\ref{gamma4}),(\ref{gamma}) and (\ref{nu}) 
the correction amplitudes appear with a negative 
sign, the critical exponents will be 
overestimated if the amplitudes are negative and they will be 
underestimated otherwise.

 Let us add finally that
 throughout  our work we have not relied solely 
on the above  numerical technique, but we also have always 
 considered various other approximations  
obtained by more conventional methods in order 
to understand, or at least to be aware of 
 any differences in the  estimates.

\section{ Results and comments} 

Since our analysis is aimed at exposing the role of the non analytic 
corrections to scaling, it is desirable  firstly to test whether the values 
of the confluent exponents  taken  from the FD perturbative  
computations are also generally consistent with the estimates,
 unfortunately not yet as precise, which can be extracted 
directly from the HT series. Indeed, as we have mentioned above, 
the amplitudes of these corrections are not universal and therefore 
 they might be negligibly small.
 One might even  suspect that our analysis is somehow artificially 
forcing on the series a behavior,  which, due to their insufficient 
 length, they are not yet able to exhibit.
 On the other hand, it has been argued that the  uncertainties 
usually quoted for the FD values of the 
renormalized couplings and of the confluent exponents might 
be unrealistically small\cite{nickel80,pv,lms,nick81}. 
In fact, one should recall that in the 
context of the three-dimensional 
 $\lambda (\vec\phi^2)^2$  field theory, the  confluent 
exponent is computed in terms of the slope 
of the beta-function at the fixed point $g_r(N)$.
As indicated in Refs.\cite{gz2,nickel80,pv}, the presence  
of non-analytic  terms, with sufficiently large amplitudes,
 in the expansion of the 
beta-function at  $g_r(N)$,  
might spoil the convergence of the estimates both of the 
renormalized couplings and of the confluent exponents. 
The ensuing uncertainties
would  reflect on the accuracy of the 
estimates of the critical exponents. Moreover
   the $g-$expansion of the critical exponents would itself be 
 directly affected by similar non-analytic contributions.
 The pragmatic point of view adopted in Ref.\cite{gz2} 
is that if these singular terms
 exist, they do not seem to have   visible effects.

Let us then show
 that the values 
of $\theta(N)$   reported  in Table 1  are 
  approximately consistent  with the actual behavior of 
the series. Assuming knowledge only of $\beta^{\#}_c(N)$, 
we have computed  the Baker-Hunter transforms\cite{bahu} 
of the $\chi$ and $\mu_2$ series   and, by  reconstructing 
 the locations and the residua of their 
 singularities, 
 we have  estimated 
exponents and amplitudes of the  critical singularity and of the leading 
 correction  to scaling. 
 Unfortunately this procedure  fails to detect  narrow and clear signals of 
 the scaling corrections for $N < 4$,  probably due  
to the small size of their amplitudes. 
But the situation is completely different
 for $N \geq 4$. In this range of values of $N$, 
  the Baker-Hunter method leads to 
 values of $\theta(N)$  fairly consistent with those  reported
 in Table 1. 
 Also the values of the correction amplitudes,
 are  compatible  
with those emerging
 from the SDA analysis to be discussed below. Moreover, the results
 are rather stable in a  relatively wide
 range of biased values for $\beta_c$.  
We regard this as 
  convincing evidence 
that the confluent corrections   cannot be 
 a by-product of our double biased analysis and 
 as a  confirmation  that their amplitudes  are not  small for $N \geq 4$.
 Unfortunately, the uncertainties which affect this method for estimating 
 the confluent exponents and the correction amplitudes 
 are still rather large.
For instance, using  the bcc lattice series for $\chi$, the Hunter-Baker
 procedure suggests
\begin{equation} 
\theta(4)=.64(4);\ \ \theta(6)=.63(4);\ \   \theta(8)=.66(4);\ \
 \theta(10)=.69(4)
\label{hunterb}\end{equation} 

A second consistency test can also be performed.
 On both lattices and for each  value 
of $N$, we have studied how our SDA estimates of $g_r(N)$ 
 depend on the biased value  used for  the confluent exponent 
 by varying it in a $20-30\%$ 
 range around the central value $\theta(N)$ indicated in Table 1.
 For all values of $N$ such that  the confluent amplitudes 
are not too small, it has been quite interesting to observe 
that, although  the estimates of $g_r(N)$ obtained from the 
sc and the bcc series 
are in general somewhat different for a generic value of $\theta$, 
they tend to become equal, 
or at least  very close, when  $\theta \simeq \theta(N)$. 

These two  tests give us further confidence
 that the main lines of this analysis and the specific 
biased values of $\theta$  used as inputs 
 are   reasonable.

\subsection{ Hyperscaling tests}

 We shall now proceed to  examine
 directly  $\chi_{4}(N,\beta)$ in order to estimate
its critical exponent  $\gamma(N)+2\Delta_4(N)$ 
 and to compare it
 with the value 
$2\gamma(N)+3\nu(N)$ it should take  if  the hyperscaling relation
 (\ref{hyp})  holds true.
 On each lattice, the analysis  has been  performed  
 by  first-order  SDA's of the effective exponent  (\ref{gamma4}) doubly  
 biased with   $\theta(N)$ and with the value of  $\beta^{\#}_c(N)$ 
 obtained in our previous (biased)  analysis\cite{bc3d} 
of the susceptibility.

  We have reported in table 2 our estimates  for 
the critical exponents of $\chi_4(N,\beta)$ 
 obtained by this procedure
together with the biased 
values of $\beta^{\#}_c(N)$ 
 and the values of  $2\gamma(N)+3\nu(N)$ obtained by the analogous biasing
procedure in our  previous analysis\cite{bc3d} of $\chi$ and $\xi^2$.
 No significant violation of universality and hyperscaling 
 is observed. Notice that no such extensive  test of 
hyperscaling exists so far in the literature. 

Let us quote a few earlier studies of this issue 
for particular values of $N$.
 In the $N=0$ case,    a study of $\chi_4$ based on (\ref{chi4saw})
  has been performed
by a Monte Carlo simulation in Ref.\cite{def}.
 The authors have measured the exponents  
$ 2\Delta_4-\gamma= 1.7317 \pm0.0074 \pm 0.0074$
 and $\nu= 0.5745 \pm 0.0087 \pm 0.0056$. 
 The final result  is 
 expressed as $3\nu+ \gamma -2\Delta_4=-0.0082 \pm 0.0027 \pm 0.018$,  
the first error being 
systematic and the second statistical.

 In the $N=1$ case, the tests of the hyperscaling relation (\ref{hyp})
 are  numerous and have a long history\cite{fisher,dombook}. The 
validity of (\ref{hyp}) for the 3d Ising model
had  been questioned  
by G. A. Baker\cite{bakeb,bake} on the basis of 
 an  analysis of  10-12 term 
series for  the sc, bcc and fcc lattices.
  A few years later, 
 when B. G. Nickel computed $O(\beta^{21})$ series on the  bcc lattice
for $\chi$ and $\mu_2$ in the spin $S$ Ising  model, 
 it became clear that rather long series were necessary 
 to allow for 
the scaling corrections and thus to obtain more satisfactory 
estimates of $\gamma$ 
and $\nu$\cite{nickel80,zinn79,adl,nr90}. On the other hand accurate 
analyses of the critical behavior of the $\chi_4(1,\beta)$ series 
to order $\beta^{17}$ on the sc lattice \cite{gutt,ns,bcpre,rehr}
 had yielded reliable values also for $\Delta_4(1)$.
On the basis of these results, as well as of various  recent Monte Carlo 
 results\cite{tsy,kl,baka}  a common consensus 
was  reached that,  for $N=1$,
 if any violation of (\ref{hyp}) occurs, it should be much smaller than
  was initially suspected.    Our contribution 
 to this issue also consists in providing an extension from 
order $\beta^{13}$ to order 
$\beta^{17}$ of the Ising bcc series for $\chi_4$, and therefore 
in further improving the accuracy of the HT test of
 hyperscaling and universality  even for the widely studied $N=1$ case.

\subsection{ Renormalized couplings}

Let us first mention that, since $\xi^2 = O(\beta)$ in the vicinity of 
$\beta=0$,
 from  the series for $\chi$, $\xi^2$ and $\chi_4$ we can form
 two distinct
auxiliary functions $w(N,\beta)$ and $u(N,\beta)$,  analytic 
at $\beta=0$, both of which, when extrapolated  at $\beta_c$
  yield $g(N,\beta_c)$ and therefore 
$g_r(N)$, if we assume the validity of 
the hyperscaling relation. More precisely we shall consider:

\begin{equation}
  u(N,\beta)\equiv 
- \frac{\xi^2(N,\beta) \chi^{4/3}(N,\beta)}{(v f(N) \chi_4 (N,\beta))^{2/3}}
\label{defb}\end{equation} 

whose value at $\beta_c(N)$ is $g_r(N)^{-2/3}$, and

\begin{equation}
  w(N,\beta)\equiv - \frac{v f(N) \chi_{4}(N,\beta)}
{\beta_c^{3/2}(\xi^2(N,\beta)/\beta)^{3/2} \chi^2(N,\beta)}
\label{defa}\end{equation} 

whose value at $\beta_c(N)$ is $g_r(N)$.

It is interesting to form approximants both of $u(\beta)$
and of $w(\beta)$ because for various values of $N$, 
 at the presently available order of expansion,  they 
 still show slightly different  convergence properties. 
This may be seen as an indication that the $\chi_{4}$ series are 
still not very long. Indeed, as we have argued 
in Ref.\cite{bcpre}, at order $\beta^s$
 the dominant contributions to the HT expansion of $\chi_4$ come 
from correlation functions of spins whose average distance is $\approx s/4$.
Therefore the presently available expansions with $s_{max}=17$ 
still describe only a rather small  system. 

 Table 3 contains our estimates of the universal 
renormalized coupling  $g_r(N)$.

For $N \leq 4$ we have evaluated $g_r(N)$  
by forming SDA's of the auxiliary 
function $w(N,\beta)$, which has been chosen 
because it yields sequences of estimates 
showing little or no residual trends 
when an increasing number of series coefficients is used.
On the other hand, for $N> 4$,  we have used 
 $u(N,\beta)$ because  the estimates  obtained  from  it
 show  the slowest (generally decreasing) 
residual   trends. Whenever relevant, we have  indicated this fact 
 by reporting asymmetric error bars.

In  the  $N=0$ case,  allowance for 
the correction to scaling yields 
 a value of $g_r(0)$  
 approximately $2\%$ smaller than the one recently obtained within the 
FD expansion\cite{gz2}, but very close to the value suggested\cite{pv} 
by the $\epsilon-$expansion. 
 Our  value 
 is also close to that indicated in Ref.\cite{mur} 
 and produces, via the seven 
loop FD perturbation series, central values  of $\gamma(0)$ and $\nu(0)$ 
$\approx 0.2\%$ lower than those  quoted in Ref.\cite{gz2}, but  within 
their error bars. 
 It is also worth recalling that also our earlier HT analysis\cite{bc3d}
 of $\chi(0,\beta)$ and $\xi^2(0,\beta)$ 
 had supported those low exponent estimates 
 in good agreement with very  recent high  precision measures by stochastic 
methods on the sc lattice\cite{causo}.

 For $N=1$, on the sc lattice, we have reported here a central estimate of 
$g_r(1)$ slightly lower than,
 though consistent with the  estimate $g_r(1)=1.411$
 obtained from our previous analysis \cite{bcpre} 
 based on  SDA's of $u(1,\beta)$, rather than of $w(1,\beta)$.

 A small sample of the most recent estimates of $g_r(1)$ 
by  various methods  has also been included in 
the table. All of them appear to be mutually consistent, if we consider 
 how difficult it has been to achieve  very accurate Monte Carlo 
measures of $g_r^{sc}(1)$\cite{tsy,kl,baka} and we recall 
  that, even in the Ising case, the previous 
 HT series estimates\cite{zlf}  of the renormalized coupling 
 were based on  expansions shorter than those presented here.
 Indeed, although  $\chi_4(1,\beta)$ on the sc lattice 
has long been known through order $\beta^{17}$, the corresponding expansion
for the renormalized coupling was not available
 before our recent work\cite{bc3d,bcpre},
 because $\xi^2(1,\beta)$ reached\cite{roskies}   only order $\beta^{15}$.
 On the other hand the $\chi_4(1,\beta)$ series for  the bcc lattice 
 was known to order $\beta^{13}$ only.

To our knowledge, no Monte Carlo results are yet available for  $N >1$. 

For $N \geq 3$ our estimates are systematically slightly higher than 
the FD values of Refs.\cite{gz2,sokoth}, and perhaps 
the residual decreasing trend
 in our estimates might not  be sufficient to 
reconcile them. This difference is related to 
 our allowance of the scaling corrections by  doubly biased
 SDA's and is consistent with the higher values of $\gamma$ and $\nu$
 that we had obtained in our  biased analysis\cite{bc3d} of 
$\chi$ and $\xi$. 
As we have stated above in discussing the general 
 features of the SDA's,   significantly larger estimates  
for $N< 4$ and somewhat lower estimates for  $N \geq 4$
 would be  obtained, if the 
renormalized couplings were evaluated by simple PA's.
 This fact is completely consistent with the  observed behavior of the 
 correction amplitudes as 
functions of $N$ to be discussed in next subsection.
 A similar observation has been made also in Ref.\cite{pv} where,
 on the basis of the old sc lattice   $O(\beta^{14})$ 
series\cite{lw88,b90}, the
$g_r(N)$  have been evaluated by ordinary DA's, either 
directly  or 
after performing a change of variable\cite{rosk,adp,nd} designed
 to regularize the leading correction to scaling and 
numerically similar to our SDA's.
 Therefore the final HT estimates  of Ref.\cite{pv}   essentially 
 agree with ours.

 We have included in table 3 some estimates of $g_r(N)$ 
 based on the $\epsilon$-expansion to order $\epsilon^4$ recently
 presented in Ref.\cite{pv}. They are  compatible with ours 
 for $N<3$, while, for $N\geq 3 $,  the central values  
 are   $\approx 2\%$ lower. 

\subsection{ Critical and correction amplitudes}

 In   tables 4  and 5 we  have reported 
 our estimates of the (non-universal) critical amplitudes $C^{sc}_{\chi}(N)$,
$C^{sc}_{\xi}(N)$ and $C^{bcc}_{\chi}(N)$,  $C^{bcc}_{\xi}(N)$,
 based on the values of $\beta_c^{\#}(N)$, 
$\gamma^{\#}(N)$ and $\nu^{\#}(N)$ obtained in the biased 
analysis of Ref.\cite{bc3d}.
Earlier determinations of the critical amplitudes 
 either from the extrapolation of 
(generally shorter) HT series or from stochastic simulations
 are also available for 
$N=1$ in Refs.\cite{zlf,rzw}, for $N=2$ in Ref.\cite{got}, 
and for $N=3$  in Ref.\cite{che}. However,
 comparisons 
with the  results of tables 4 and 5, which in general 
are  close to the earlier ones, 
are not very illuminating, since  the 
estimates depend sensitively on the numerical procedures,
 on the biased values used for $\beta^{\#}_c(N) $ 
and on the relevant critical exponents, which are slightly different in the 
various studies. 

For example in the $N=1$ case\cite{liu1,zlf} 
(on the basis of shorter 
sc lattice series, but the same bcc series), 
the following estimates are proposed:
$C^{sc}_{\chi}(1)=1.0928(10)$ and $C^{sc}_{\xi}(1)=0.4984^{(10)}_{(50)}$;  
$C^{bcc}_{\chi}(1)=1.0216(8)$ and $C^{bcc}_{\xi}(1)=0.4608(2)$
assuming $\beta_c^{sc}(1)=0.221630(12)$,
$\beta_c^{bcc}(1)=0.157368(7)$, $\gamma(1)=1.2395$ and $\nu(1)=0.632$. 

For $N=2$, in Ref.\cite{got}  the estimates  $C^{sc}_{\chi}(2)=1.058(7)$ 
 and $C^{sc}_{\xi}(2)=0.498(2)$ have been obtained from a fit to 
Monte Carlo data, assuming  $\gamma(2)=1.3160(25)$, $\nu(2)=0.669(2)$, 
allowing for confluent corrections with exponent  
 $\theta(2)=0.522$.
 This fit also yields $\beta_c^{sc}(2)=0.454162(9)$
 from the analysis of $\chi$, and
$\beta_c^{sc}(2)=0.454167(10)$, 
 from the analysis of $\xi$. 

For $N=3$, in Ref.\cite{che},  the estimates  $C^{sc}_{\chi}(3)=0.955(6)$ 
 and $C^{sc}_{\xi}(3)=0.484(2)$ have been obtained from a fit to 
Monte Carlo data (with no allowance for confluent corrections)
  also yielding $\beta_c^{sc}(3)=0.69294(3)$,
 $\gamma(3)=1.391(3)$
 from the analysis of $\chi$ and
$\beta_c^{sc}(3)=0.69281(4)$, $\nu(3)=0.698(2)$,
 from the analysis of $\xi$. 

As  has been stressed in general
 in Ref.\cite{liufi} and as we have anticipated 
in our considerations of Section 3  
 on the numerical properties of SDA's, the discussion
 of the estimates of the scaling correction amplitudes is much more delicate.
 Let us first comment on some qualitative features of 
 the estimates  of these amplitudes for  the sc and the bcc lattices 
 which are   denoted as 
 $a^{sc}_{\chi}(N)$, $a^{bcc}_{\chi}(N)$,
  $a^{sc}_{\xi}(N)$, $a^{bcc}_{\xi}(N)$, $a^{sc}_{g}(N)$, $a^{bcc}_{g}(N)$,
 and reported in table 6.

 Both correction amplitudes $a^{\#}_{\chi}(N)$ and $a^{\#}_{\xi}(N)$ 
are negative for $N \lesssim 2$,  whereas they are positive 
 and increasing for $N>2$. (Actually we have reported a positive  value 
 for $a^{bcc}_{\chi}(2)$, but
 with a large uncertainty.)
Therefore the ratio $a_{\xi}/a_{\chi}$ is very likely 
to be positive for all values of $N$.  The  correction amplitudes 
 $a^{\#}_{\chi}(N)$ and $a^{\#}_{\xi}(N)$  turn out to be
 small, but not negligible 
  for $N \leq 1$ and rather large for $N \geq 4$, both  in the sc 
 and in the bcc lattice case. On the contrary, 
for $N=2$ and $N=3$ they  are very small. 
 Thus the overall behavior of the  
 correction amplitudes for $\chi$ and $\xi$ as functions of $N$ 
 appears to be smooth 
 and  completely consistent with the  size and the sign of 
the differences\cite{bc3d} between our unbiased estimates of the 
critical exponents $\gamma$ and $\nu$  and the corresponding 
estimates biased  with 
both $\beta_c$ and $\theta$. More precisely, 
 we recall that  the 
non-analytic corrections to scaling lead to 
 (slightly) higher effective exponents for $N<2$ and 
 to (significantly) lower effective exponents for 
$N \geq 4$. 
On the other hand, the $a^{\#}_{g}(N)'s$ are positive
 for $N \lesssim 2$, while they are negative
 and decreasing for $N>2$, so that the ratio $a_{g}/a_{\chi}$ is
  negative for any $N$. (Actually we have reported a positive  value 
 for $a^{sc}_{g}(3)$, but
 with  a large uncertainty.) 
The $a^{\#}_{g}(N)'s$ are generally not small,
 except for  $N=3$ in the sc case and for  $N=2$ in the bcc case.

It is appropriate now to quote some earlier  
evaluations of  $a_{\chi}$  and  $a_{\xi}$ by HT series 
or Monte Carlo simulations. 
In the $N=1$ case, it had been established\cite{fisher,nr90,liufi,george}
 long ago that the sign of  $a_{\chi}(1)$ and 
of  $a_{\xi}(1)$ is negative 
 on the sc,  bcc  and  fcc lattices. 
In Ref.\cite{nr90}, for the spin 1/2 Ising model 
on the bcc lattice,  the estimates  
$a^{bcc}_{\chi}(1)\approx -0.13$ and $a^{bcc}_{\xi}(1)\approx -0.11$
 have been indicated, together with the central values 
$\gamma(1)=1.237$, $\nu(1)=0.630$ and $\theta(1)=0.52$, 
on the basis of a second order DA analysis. 
For $N=2$, the above cited Monte Carlo simulation of Ref.\cite{got}, 
  yielded the estimates  $a^{sc}_{\chi}(2)=-0.15(6)$ 
 and $a^{sc}_{\xi}(2)=-0.20(4)$. 
 Clearly, in both cases the critical parameters 
are slightly different from ours 
and this is sufficient to explain the 
somewhat different estimates for the correction amplitudes.

In the spin 1/2 Ising  case, it has been 
argued long ago\cite{nickel80} that 
$a_{g}^{bcc}(1)$ should be large. Recently\cite{pv},
 it also has  been observed that, if the sc lattice  Monte Carlo data of 
Ref.\cite{baka} are simply fitted by the function   
$g_r(\beta)=g^{*}(1+a_g\tau^{1/2})$,
 the value $a_{g}^{sc}(1) \approx 1.13$ is obtained 
in fair agreement with our own estimate. 

In table 7 and 8  we have listed some earlier 
 estimates of the universal ratios 
$a_{\xi}(N)/a_{\chi}(N)$  and   $a_g(N)/a_{\chi}(N)$
of  correction amplitudes obtained by various 
methods\cite{mur,zinn,nr90,rehr,ch,cr,bb}. 
 We believe that, for   $N < 4 $, it is not very  meaningful 
to quote 
 the ratios  of our central estimates  of $a_{\xi}$, $a_{\chi}$ 
 and $a_{g}$. Indeed,  
as we have already pointed out, for these values of $N$,  the amplitudes 
$a_{\xi}$, $a_{\chi}$  are small and  very sensitive to 
 the biased inputs. As a consequence, these parameters must be finely tuned,
 which cannot be justified 
 until longer series will be computed. 
 We shall indicate below a possible alternative way out of this difficulty.
 However, the case $N=1$  deserves further comment. In this case, 
on the sc lattice, a very accurate\cite{blh} determination 
of $\beta_c^{sc}(1)$ is available, and also the value 
of $\beta_c^{bcc}(1)$\cite{nr90} appears to be sufficiently  safe, so that
 the ratios of our central estimates of the amplitudes are more 
trustworthy and we have reported them in parentheses. 
 
It is interesting  to recall also that, for $N=1$, 
 suggestions that $a_g/a_{\chi}$ should be large came both from earlier 
HT estimates\cite{george}  on the fcc lattice 
($a_g^{fcc}(1)/a_{\chi}^{fcc}(1)\approx 3.9$) and from the RG estimates 
 reported in Table 8. This is a further hint that  the 
corrections to scaling should not be neglected in computing $g_r(1)$. 

 Unfortunately, the $\epsilon$-expansions of these universal 
ratios presently only  
extend to second order\cite{ch,cr},   so that  
 again we have to  point out that the uncertainty 
of the corresponding  estimates 
  might  be larger than indicated. 
 As we have mentioned above, even the estimates of these ratios from the 
 much longer FD expansions\cite{mur,bb} might have problems.
For $N <4$,  as already observed,  all series including ours 
 are   too short to accurately extract 
 the correction amplitudes. This is particularly the case for $\chi_4$. 
 Moreover, when longer series  become available, 
   our approximation procedures  might need  some improvement.
 Nevertheless these first   results from HT series 
on an extended range of values of $N$ 
seem to be  qualitatively very    reasonable. 

 As an indication of work in progress, we wish to add 
 that, even within the present order of expansion, 
 somewhat more accurate estimates of the critical parameters 
 are likely to  be obtained  by proceeding systematically 
 in the spirit of the Chen, Fisher, Nickel and Rehr approach\cite{fisher,nr90}.
 In the $N=1$ case on the bcc lattice, these authors have examined 
 HT series for   families of models 
specified by an appropriate continuous auxiliary parameter.
 The members of these families interpolate between the spin 1/2 Ising and 
the Gaussian model and
 all of them  are good candidates for belonging 
to the same universality class. 
 (This approach  easily generalizes, 
in various ways, to other values of $N$ and it is a virtue of 
 the LCE method that the corresponding series 
can be derived essentially with no  further computational effort.)
By varying the auxiliary parameter, these authors have selected 
 representative 
models such that  the leading  correction 
amplitudes $a_{\xi}$, $a_{\chi}$ (and $a_{g}$) vanish. 
 It is clear that, under these conditions,  
even by employing ordinary unbiased DA's,  the accuracy  
of the estimates of the  universal quantities can be improved dramatically.  
On the other hand, within the same approach, it is probable that
 also  the correction amplitudes 
 will be more  accurately  measured
by focusing on representative models in which   they are sufficiently 
large, provided, of course, that the subleading terms are not even larger.
 Thus  more reliable estimates 
might be achieved for their universal ratios, in particular in the range $N<4$.

\section{ Conclusions} 

The main result of this paper is the 
 extension  through $O(\beta^{17})$ of the series for $\chi_4(N,\beta)$,
 for  arbitrary $N$,  
 on the sc and on the bcc lattices. 
 Both sets  of expansion coefficients have been tabulated in the 
 appendix in order to make independent checks 
of their correctness  and alternative  analyses conveniently feasible.

 A second interesting result 
is the numerical analysis of the critical behavior 
of $\chi_4(N,\beta)$  which confirms fairly well   the validity of 
 universality and  hyperscaling  over   a wide range of values of $N$.
 We have also presented a first estimate of the size of the 
  scaling corrections   for $\chi$, $\xi^2$ and $\chi_4$
 and,  allowing for them, we have  improved the accuracy 
in the determination of the critical amplitudes and of the 
renormalized couplings.

 The agreement between our estimates of $g_r(N)$ and those from the RG 
approaches is generally fair, but not always perfect.
At this level of approximation, it is premature to emphasize
 such minor discrepancies. 
 We    believe, however, that 
longer HT series for all quantities studied here
  and perhaps improved analyses
 are still of some interest   to  achieve more reliable
 estimates and to reduce 
  the error bars substantially. Considering the performance of our codes, 
these are presently
 quite realistic objectives and, therefore, work is presently in progress to 
compute  further expansion coefficients.

\acknowledgments 
This work has been partially supported by MURST. We thank 
Prof. A. J. Guttmann
for critically reading the first draft of this paper and
 Prof. A. I. Sokolov for making available to us before 
publication his  results on the RG perturbative computation 
of the confluent exponents.

\begin{table}
\caption{ Values for $\theta$ used in our biased evaluations 
 and  determined by FD perturbative expansion} 
\label{tabella1}
\begin{tabular}{ccccccccc}
N  &0 &1 &2 &3 &4 &6 &8  &10  \\
\tableline
$\theta$&.478(10)  &.504(8) &.529(8)&.553(12)&.573(20)&.626(10)&.670(10)& 
.707(10)\\
\end{tabular}
\end{table}

\begin{table}
\caption{Verification of  hyperscaling for various values of $N$.}
\label{tabella2}
\begin{tabular}{rllll}
N &Lattice  &$\beta_c$\cite{bc3d} &  $\gamma+ 2\Delta$ 
 &$2\gamma+3\nu$\cite{bc3d}\\
\tableline
0& sc & .213493(3) &    4.10(2)      &4.0822(34) \\
 & bcc  &.153128(3) &    4.081(8)      &4.0801(34)\\
\tableline
1 & sc  &.2216544(3)\cite{blh}          & 4.361(8)     &4.3721(44) \\
 & bcc  &.157373(2)\cite{nr90}          &4.366(6)      &4.3692(27)  \\
\tableline
2 & sc  &.45419(3)  &    4.665(20)      &4.675(12)     \\
  & bcc  &.320427(3)          & 4.663(15)     &4.666(12)          \\
\tableline
3 & sc  &.69305(4)  &     4.953(20)     &4.960(12)           \\
  & bcc  &.486820(4)          &4.948(15)      & 4.946(12)   \\
\tableline
4 & sc  &.93600(4)  &     5.24(2)     &5.259(17)          \\
  & bcc  &.65542(3)          & 5.22(2)     & 5.236(17)   \\
\tableline
6 & sc  &1.42895(6)            & 5.67(2)        &5.691(19)  \\
  & bcc  &.99644(4)              & 5.65(2)            &5.673(17) \\
\end{tabular}
\end{table}

\begin{table}
\caption{  The renormalized coupling constant $g_r(N)$ for a range of 
values of $N$ on the sc and the bcc lattice as obtained by various methods.}
\label{tabella3}
\begin{tabular}{clllll}
 $N$ & HT sc & HT bcc & $\epsilon$-exp.
 & $FD$  exp. &  Monte Carlo\\
\tableline
0&1.388(5)&1.387(5)&1.390(17)\cite{pv} &1.413(6)\cite{gz2}& \\
0&       &     &                      &1.39 \cite{mur}& \\
1&1.408(7)&1.407(6)&1.397(8)\cite{pv}&1.411(4)\cite{gz2}&1.391(30)\cite{tsy}\\
1&1.459(9)\cite{zlf} &     &        &1.40 \cite{mur}&1.462(12)\cite{kl}\\
2&1.411(8)  &1.411(6)&1.413(13)\cite{pv} &1.403(3)\cite{gz2}& \\
2& &     &                   &1.40 \cite{mur}& \\
3&1.409(10)&1.406(8)&1.387(7)\cite{pv} &1.391(4)\cite{gz2}& \\
3&       &       &                      &1.39 \cite{mur}& \\
4&1.392(10)&1.394(10)&1.366(15)\cite{pv} &1.377(5)\cite{gz2}& \\ 
4&       &       &                      &1.3745\cite{sokoth}& \\ 
6&$1.355^{(+5)}_{(-10)}$  &$1.360^{(+5)}_{(-10)}$& &1.3385\cite{sokoth}& \\
8&$1.320^{(+8)}_{(-15)}$&$1.326^{(+8)}_{(-15)}$
&1.295(7)\cite{pv}&1.3045\cite{sokoth}& \\
 10 &$1.290^{(+8)}_{(-15)}$&$1.294^{(+8)}_{(-15)}$&&1.2745\cite{sokoth}& \\
\end{tabular}
\end{table}

\begin{table}
\caption{   Critical amplitudes 
on the sc  lattice for various values of $N$.}
\label{tabella4}
\begin{tabular}{llllll}
N & $\beta_c(N)$&$\gamma(N)$ &$\nu(N)$&
$C^{sc}_{\chi}(N)$&$C^{sc}_{\xi}(N)$\\
\tableline
0&0.213493(3) &1.1594(8)&0.5878(6) &1.115(1) &0.5101(3) \\
1&0.2216544(3) &1.2388(10)&0.6315(8) &1.111(1) &0.5027(3) \\
2&0.45419(3) &1.325(3)&0.675(2) &1.014(1) &0.4814(3)  \\
3&0.69305(4) &1.406(3)&0.716(2) &0.9030(8) &0.4541(2) \\
4&0.93600(4)  &1.491(4) &0.759(3) &0.7571(8) &0.4155(2) \\
6&1.42895(6) &1.614(5)&0.821(3) &0.6054(8) &0.3708(2)  \\
$\infty$& &2. &1. &0.392287..\cite{joy} &0.314870..\cite{joy}\\
\end{tabular}
\end{table}

\begin{table}
\caption{   Critical amplitudes 
on the  bcc lattice for various values of $N$.}
\label{tabella5}
\begin{tabular}{llllll}
N & $\beta_c(N)$&$\gamma(N)$ &$\nu(N)$&$C^{bcc}_{\chi}(N)$ &
$C^{bcc}_{\xi}(N)$\\
\tableline
0 &0.153128(3) &1.1582(8)&0.5879(6) &1.087(1) &0.4846(2) \\
1 &0.157373(2) &1.2384(6)&0.6308(5) &1.034(1) &0.4659(2) \\
2 &0.320427(3) &1.322(3)&0.674(2) &0.918(1) &0.4371(2)  \\
3 &0.486820(4) &1.402(3)&0.714(2) &0.794(1) &0.4072(2)  \\
4 &0.65542(3) &1.484(4)&0.756(3) &0.6580(8) &0.3691(2)  \\
6 &0.99644(4) &1.608(4)&0.819(3)&0.5020(6) &0.3231(2)  \\
$\infty$& &2. &1. &0.299741..\cite{joy}&0.263818..\cite{joy}\\
\end{tabular}
\end{table}

\begin{table}
\caption{  Correction  amplitudes 
on the sc and the bcc lattices for various values of $N$.}
\label{tabella6}
\begin{tabular}{lllllll}
N &$a^{sc}_{\chi}(N)$&$a^{bcc}_{\chi}(N)$ &$a^{sc}_{\xi}(N)$&
$a^{bcc}_{\xi}(N)$& $a^{sc}_{g}(N)$&$a^{bcc}_{g}(N)$\\
\tableline
0 &-0.022(10)&-0.05(3)&-0.11(3) &-0.1   &1.5(3)    &1.9(4)  \\
1 &-0.10(3)&-0.08(3)&-0.12(3)  &-0.08(3)  &1.0(2)    &1.0(2)  \\
2 &-0.04(2)&0.01(2) &-0.07(3) &-0.005(9) &0.39(8)    &0.14(3)  \\
3 &0.06(3) &0.17(3) &0.003(6)  &0.09(3)   &0.05(10)    &-0.29(6)  \\
4 &0.30(6) &0.5(1) &0.14(3)   &0.28(6)   &-0.12(3)    &-0.55(10)  \\
6 &0.73(15) &1.1(2) &0.37(8)   &0.60(15)   &-0.35(8)    &-0.77(20)  \\
8 &1.1(2) &1.8(3) &0.53(10)   &0.92(20)   &-0.43(10)    &-1.0(2)  \\
\end{tabular}
\end{table}

\begin{table}
\caption{  The universal ratios of correction amplitudes
 $a_{\xi}(N)/a_{\chi}(N)$ for various values of $N$. }
\label{tabella7}
\begin{tabular}{lllll}
  $N$ & HT sc & HT bcc & $\epsilon$-exp.
 & $FD$  exp. \\
\tableline
  0  &    &      &  &0.885(50)\cite{mur}           \\
 1&\Big(1.2(4)\Big)&$\Big(1.0(4)\Big)$ &0.65\cite{ch}&0.762(30)\cite{mur} \\
  1 &    &0.70(3)\cite{george};0.85(5)\cite{nr90}&0.56(15)\cite{zinn}
 & 0.65(5)\cite{bb}          \\
  2 & &   & 0.63\cite{ch}     &0.687(10)\cite{mur} \\
  2 &    &      &    & 0.615(5)\cite{bb}         \\
  3 &    &      &          &0.637(5)\cite{mur}  \\
  3 &    &      &          &0.60\cite{bb}  \\
  4 &0.49(15) &0.55(15)      &          & \\ 
  6 &0.51(15) &0.55(15)      &          & \\
  8 &0.49(15) &0.54(15)      &          & \\
\end{tabular}
\end{table}

\begin{table}
\caption{  The universal ratios of correction amplitudes 
 $a_{g}(N)/a_{\chi}(N)$ for various 
values of $N$. }

\label{tabella8}
\begin{tabular}{cccccc}
 $N$ & HT sc & HT bcc & $\epsilon$-exp.
 & $FD$  exp. \\
\tableline
  0     &    &      &         &                          \\
  1 &$\Big(-3.0(5)\Big)$ &$\Big(-3.5(5)\Big)$& -2.2(5)\cite{rehr}
&-2.85(6)\cite{bb}  \\
  2     &    &      &          &       -2.08(5)\cite{bb}  \\
  3     &    &      &          &-1.65(4)\cite{bb} \\
  4     &    &      &          & \\ 
  6     &-1.2(4) &-0.7(2)      &          & \\
  8     &-1.1(4) &0.8(3)      &          & \\
\end{tabular}
\end{table}

\appendix
 
\section{The second field derivative of the susceptibility on the sc lattice}

The HT expansion coefficients of the 
second field derivative of the susceptibility

$  \chi_{4}(N,\beta )={\frac {3N} {N+2}}  \sum_{x,y,z}
\langle  v(0) \cdot v(x) v(y) \cdot v(z)\rangle_{c}=
{\frac {3N} {N+2}} \Big(-\frac {2} {N} 
+ \sum_{r=1}^\infty d_r(N) \beta^r \Big) $ 

 on the sc lattice are

\scriptsize

\[d_1(N)= -48/{N^2}\]

\[d_2(N)=\frac{ -1248 - 660\,N} { {N^3}\,\left( 2 + N \right) }\]

\[d_3(N)=\frac{-12480 - 6912\,N} {{N^4}\,\left( 2 + N \right)}\]

\[d_4(N)=\frac{  -851712 - 1128192\,N - 474000\,{N^2} - 61236\,{N^3}}
{ {N^5}\,{{\left( 2 + N \right) }^2}\,\left( 4 + N \right) }\]

\[d_5(N)=\frac{   -6573312 - 8786880\,N - 3725856\,{N^2} - 483840\,{N^3}}
{  {N^6}\,{{\left( 2 + N \right) }^2}\,\left( 4 + N \right) }\]

\[d_6(N)=\frac{   -565908480 - 1137490944\,N - 877991616\,{N^2} 
- 321566208\,{N^3} -    55124016\,{N^4} - 3514968\,{N^5}}
{  {N^7}\,{{\left( 2 + N \right) }^3}\,\left( 4 + N \right) \,
   \left( 6 + N \right)}\]

\normalsize

For the coefficients which follow it is typographically
more convenient to set $d_r(N)=P_r(N)/Q_r(N)$ and to
 tabulate separately the numerator polynomial $P_r(N)$
 and the denominator polynomial $Q_r(N)$,
\scriptsize

\[P_7(N)=  -3849744384 - 7739114496\,N 
 - 5976661248\,{N^2} - 2190260352\,{N^3} - 
   375495552\,{N^4} - 23938560\,{N^5}\]

\[Q_7 (N)=   {N^8}\,{{\left( 2 + N \right) }^3}\,
   \left( 4 + N \right) \,\left( 6 + N \right) \]

\[P_8(N)=  -1607361822720 - 4630934495232\,N - 5658731108352\,{N^2} - 
   3815032300032\,{N^3} - 1545703906176\,{N^4}\]\[ - 383951922240\,{N^5} - 
   56942341632\,{N^6} - 4600824312\,{N^7} - 154867284\,{N^8}\]

\[Q_8 (N)= {N^9}\,{{\left( 2 + N \right) }^4}\,{{\left( 4 + N \right) }^2}\,
   \left( 6 + N \right) \,\left( 8 + N \right) \]

\[P_9(N)=  -10146634334208 - 29145299066880\,N - 35515117682688\,{N^2} - 
   23883563143680\,{N^3} - 9654774400512\,{N^4}\]\[ - 2393329321728\,{N^5} - 
   354292592640\,{N^6} - 28580124768\,{N^7} - 960719616\,{N^8}\]

\[Q_9 (N)= {N^{10}}\,{{\left( 2 + N \right) }^4}\,
{{\left( 4 + N \right) }^2}\,
   \left( 6 + N \right) \,\left( 8 + N \right)\]

\[P_{10}(N)=  -4986778413957120 - 18502905604472832\,N 
 - 30434129019666432\,{N^2} - 
   29229711376023552\,{N^3} \]\[- 18173047179460608\,{N^4} - 
   7663189901412864\,{N^5} - 2231748901824768\,{N^6} - 
   448039233434880\,{N^7}\]\[
   - 60661040264064\,{N^8}
	 - 5267106682272\,{N^9} - 
   263609235360\,{N^{10}} - 5754914568\,{N^{11}},\]

\[Q_{10}(N)={N^{11}}\,{{\left( 2 + N \right) }^5}\,
{{\left( 4 + N \right) }^3}\,
   \left( 6 + N \right) \,\left( 8 + N \right) \,\left( 10 + N \right)\]

\[P_{11}(N)=  -29957292231229440 - 110697821461807104\,N 
 - 181359443339575296\,{N^2} - 
   173526582721855488\,{N^3}\]\[ - 107505154356572160\,{N^4} - 
   45183600633858048\,{N^5} - 13119088167641088\,{N^6} - 
   2626537796155392\,{N^7}\]\[ - 354741723247872\,{N^8}
 - 30735239578752\,{N^9} - 
   1535369868288\,{N^{10}} - 33465664512\,{N^{11}}\]

\[Q_{11}(N)={N^{12}}\,{{\left( 2 + N \right) }^5}\,{{\left( 4 
 + N \right) }^3}\,\left( 6 + N \right) \,\left( 8 
 + N \right) \,\left( 10 + N \right)\]

\[P_{12}(N)=  -25424963775485706240 - 112568160566969892864\,N - 
   226122593284806672384\,{N^2} - 272739288108751650816\,{N^3}\]\[ - 
   220335024310830366720\,{N^4} - 125926002861175824384\,{N^5} - 
   52429589825842464768\,{N^6} - 16133360608103531520\,{N^7}\]\[ - 
   3682631735427354624\,{N^8} - 619886922488017920\,{N^9} - 
   75677317494258048\,{N^{10}} - 6492816409650048\,{N^{11}}\]\[ - 
   369850295426784\,{N^{12}} - 12514445149200\,{N^{13}} - 
   189708636600\,{N^{14}}\]

\[Q_{12}(N)={N^{13}}\,{{\left( 2 + N \right) }^6}\,
{{\left( 4 + N \right) }^3}\,
   {{\left( 6 + N \right) }^2}\,\left( 8 + N \right) \,
   \left( 10 + N \right) \,\left( 12 + N \right) \]

\[P_{13}(N)=-147436255220890337280 - 649808498848293715968\,N - 
   1299512129528339103744\,{N^2} \]\[- 1560658843095801004032\,{N^3} - 
   1255571803469692796928\,{N^4} - 714760376636548620288\,{N^5}\]\[ - 
   296490805364682670080\,{N^6} - 90920656735938183168\,{N^7} - 
   20688048472922093568\,{N^8} \]\[- 3472336842523894272\,{N^9} - 
   422813491673485824\,{N^{10}} - 36192084771724800\,{N^{11}}\]\[ - 
   2057419542670080\,{N^{12}} - 69492377025984\,{N^{13}} - 
   1051829121024\,{N^{14}}\]

\[Q_{13}(N)={N^{14}}\,{{\left( 2 + N \right) }^6}\,
{{\left( 4 + N \right) }^3}\,
   {{\left( 6 + N \right) }^2}\,\left( 8 + N \right) \,
   \left( 10 + N \right) \,\left( 12 + N \right) \]

 \[P_{14}(N)=-566463587862368653148160 - 
   3043875242905724409348096\,N - 7576777562322913437155328\,{N^2}\]\[ - 
   11598723404301679271608320\,{N^3} - 12226130280460910006894592\,{N^4} - 
   9415725334597698602139648\,{N^5}\]\[ - 5486072371880331913592832\,{N^6} - 
   2470783134534602565992448\,{N^7} - 871335330057590064906240\,{N^8}\]\[
 -    242243678099334349271040\,{N^9} - 53184931154543324258304\,{N^{10}} - 
   9193953185463740998656\,{N^{11}}\]\[
   - 1241394935192535991296\,{N^{12}} - 
   129075746454058556160\,{N^{13}} - 10102244514482628864\,{N^{14}}\]\[ - 
   573981185124034560\,{N^{15}} - 22283021804865984\,{N^{16}}
 -  527221923353952\,{N^{17}} - 5719330613520\,{N^{18}}\]

\[Q_{14}(N)={N^{15}}\,{{\left( 2 + N \right) }^7}\,
{{\left( 4 + N \right) }^4}\,
   {{\left( 6 + N \right) }^3}\,\left( 8 + N \right) \,
   \left( 10 + N \right) \,\left( 12 + N \right) \,\left( 14 + N \right)\]

\[P_{15}(N)=  -3199047998804219563868160 - 17113814040741606825394176\,N - 
   42413540836361700246552576\,{N^2}\]\[ - 64650199580585358413266944\,{N^3} - 
   67863921600679932441133056\,{N^4} - 52054127611661734870253568\,{N^5}\]\[ - 
   30212418901698609794777088\,{N^6} - 13556920252810313722822656\,{N^7} - 
   4764324172851484585525248\,{N^8}\]\[ - 1320233119260810969808896\,{N^9} - 
  288978328415607033544704\,{N^{10}} - 49814490389339231121408\,{N^{11}}\]\[ - 
  6708703760313406396416\,{N^{12}} - 695898456216849782784\,{N^{13}} - 
   54348379646212915200\,{N^{14}}\]\[ - 3081948095561442816\,{N^{15}} - 
   119439454925014272\,{N^{16}} - 2821599683684352\,{N^{17}} - 
   30566943406080\,{N^{18}}\]

\[Q_{15}(N)={N^{16}}\,{{\left( 2 + N \right) }^7}\,
{{\left( 4 + N \right) }^4}\,
   {{\left( 6 + N \right) }^3}\,\left( 8 + N \right) \,
   \left( 10 + N \right) \,\left( 12 + N \right) \,\left( 14 + N \right)\]

\[P_{16}(N)= -18299002328234221207226941440 - 
   114585264379209072949397028864\,N - 
   336677952680726802363970486272\,{N^2}\]\[ - 
   617145013898356017378109685760\,{N^3} - 
   791521185299488842160294330368\,{N^4}\]\[ - 
   755125087023410853999229796352\,{N^5} - 
   556163233950005337533492232192\,{N^6}\]\[ - 
   323972815609871396650371514368\,{N^7} - 
   151693834060073149464108859392\,{N^8}\]\[ - 
   57712428469432137886012145664\,{N^9} - 
   17963030647495383412208861184\,{N^{10}}\]\[ - 
   4590796640925368216011948032\,{N^{11}} - 
   964233613788142396008136704\,{N^{12}}\]\[ - 
   166150396381094082218360832\,{N^{13}} - 
   23381941196613066788198400\,{N^{14}}\]\[ - 
   2666695999548097875723264\,{N^{15}} 
 - 243616432556930571577344\,{N^{16}}\]\[ - 
   17525072840959462713984\,{N^{17}} - 968152139098787129664\,{N^{18}} - 
   39538186077707686464\,{N^{19}} \]\[- 1121781497993998176\,{N^{20}} - 
   19697772998733576\,{N^{21}} - 160868281485204\,{N^{22}}\]

\[Q_{16}(N)=    {N^{17}}\,{{\left( 2 + N \right) }^8}\,{{\left( 4 
+ N \right) }^5}\, {{\left( 6 + N \right) }^3}\,{{\left( 8 + N \right) }^2}\,
   \left( 10 + N \right) \,\left( 12 + N \right) \,\left( 14 + N \right) \,
   \left( 16 + N \right)\]

\[P_{17}(N)=  -101235958788462673253105664000 
 - 631215419419149274713371443200\,N - 
   1846824461720939657841963171840\,{N^2}\]\[ - 
   3371230360792179922631399571456\,{N^3} - 
   4306156546428529311715461955584\,{N^4}\]\[ - 
   4091812122737075444969361113088\,{N^5} - 
   3002060652157587317256634761216\,{N^6}\]\[ - 
   1742224189122069022527046287360\,{N^7} - 
   812841546356917149508011294720\,{N^8}\]\[ - 
   308190677384300600694770368512\,{N^9} - 
   95612813294752704503595663360\,{N^{10}}\]\[ - 
   24360624222950336197451317248\,{N^{11}} - 
   5101833629477951341726236672\,{N^{12}}\]\[ - 
   876738435402445320757764096\,{N^{13}} - 
   123070791802592404991434752\,{N^{14}}\]\[ - 
   14003418390237375888211968\,{N^{15}} - 
   1276532972320935406141440\,{N^{16}} \]\[
- 91648782429013822778880\,{N^{17}}- 
   5053887369689137387008\,{N^{18}} - 206055475899900727296\,{N^{19}}\]\[ - 
   5837513893110173568\,{N^{20}} - 102365100472039776\,{N^{21}} - 
   834986039880960\,{N^{22}}\]

\[Q_{16}(N)=  {N^{18}}\,{{\left( 2 + N \right) }^8}\,
   {{\left( 4 + N \right) }^5}\,{{\left( 6 + N \right) }^3}\,
   {{\left( 8 + N \right) }^2}\,\left( 10 + N \right) \,
   \left( 12 + N \right) 
 \,\left( 14 + N \right) \,\left( 16 + N \right) \]

\normalsize

	In particular for $N=0$
	we have (in terms of the variable 
	$\tilde \beta=\beta/N$) 

\scriptsize

\[\hat \chi_4(\tilde\beta)=       -3
                               -72 \tilde\beta
                              -936 \tilde\beta^{ 2} 
                             -9360 \tilde\beta^{ 3} 
                            -79848 \tilde\beta^{ 4} 
                           -616248 \tilde\beta^{ 5} 
                          -4421160 \tilde\beta^{ 6} 
                         -30076128 \tilde\beta^{ 7} 
                        -196211160 \tilde\beta^{ 8} \]\[
                       -1238602824 \tilde\beta^{ 9} 
                       -7609219992 \tilde\beta^{10} 
                      -45711200304 \tilde\beta^{11} 
                     -269412610536 \tilde\beta^{12} 
                    -1562290776792 \tilde\beta^{13} 
                    -8932238341992 \tilde\beta^{14} \]\[
                   -50443946980992 \tilde\beta^{15} 
                  -281783630311272 \tilde\beta^{16} 
                 -1558917555928200 \tilde\beta^{17}  \]

\normalsize

	For $N=1$ [ the spin 1/2 Ising model], we have 
\scriptsize
\[ \chi_4(1,\beta)=-2 -48\beta -636\beta^{2 } -6464\beta^{ 3}
 -55892\beta^{ 4} -2174432/5\beta^{ 5} -47009464/15\beta^{ 6}
 -2239468288/105\beta^{ 7} 
  -14570710772/105\beta^{ 8}\]\[ -823130010272/945\beta^{ 9} 
-25080975789304/4725\beta^{ 10} 
  -1640401398782848/51975\beta^{ 11}
 -28654566671774104/155925\beta^{ 12} \]\[
  -2130434175575247424/2027025\beta^{ 13} 
-83969257269976828688/14189175\beta^{ 14} 
  -6995762565293277161216/212837625\beta^{ 15}\]\[
 -38389375874347206695732/212837625\beta^{16 } 
  -272537955948789968719904/278326125\beta^{17}..\]

\normalsize

	For $N=2$ [the XY model] we have 
\scriptsize
\[ \chi_4(2,\beta)=-3/2 -18\beta -963/8\beta^{2 } -1233/2\beta^{ 3}
 -171687/64\beta^{ 4} -167661/16\beta^{ 5} -38749413/1024\beta^{ 6} 
  -32973957/256\beta^{ 7}\]\[ -2142639141/5120\beta^{ 8}
 -13411622379/10240\beta^{ 9} -3907085119879/983040\beta^{ 10} 
  -240713424017/20480\beta^{ 11}\]\[ -1247905418479081/36700160
\beta^{ 12} -166057186013983/1720320\beta^{ 13} \]\[
  -407002859073704999/1509949440\beta^{14 }
 -1960425200264079271/2642411520\beta^{ 15} \]\[
  -3835682132124206551811/1902536294400\beta^{ 16}
 -245360122597207497559/45298483200\beta^{ 17}...\]

\normalsize

	For $N=3$ [the Heisenberg classical model], we have 
\scriptsize

\[ \chi_4(3,\beta)=-6/5 -48/5\beta -1076/25\beta^{2 }
 -11072/75\beta^{ 3} -677044/1575\beta^{ 4} -981856/875\beta^{ 5} 
  -958584296/354375\beta^{ 6}\]\[ -261075968/42525\beta^{ 7}
 -362572843588/27286875\beta^{ 8} 
  -9268612328224/334884375\beta^{ 9}\]\[
 -3647348945492264/65302453125\beta^{ 10} 
  -1655707479102099328/15084866671875\beta^{ 11} \]\[
 -3670221101428789064/17405615390625\beta^{ 12} 
  -53935317813474946624/135763800046875\beta^{ 13}  \]\[
  -99068754350666844524336/134632435046484375\beta^{ 14}  
  -4880947680478330092600064/3635075746255078125\beta^{ 15}  \]\[
  -192879202499123356385626829692/79771737251567689453125\beta^{ 16}\]\[  
  -38137398242459901685390609504/8863526361285298828125\beta^{17 }...\]

\normalsize

\section{The second field derivative of 
the susceptibility on the bcc lattice}

The HT expansion coefficients of the 
second field derivative of the susceptibility
 on the bcc lattice are

\scriptsize

\[d_1(N)= \frac{ -64}{{N^2}}\]

\[d_2(N)= \frac{-2304 - 1200\,N} {{N^3}\,\left( 2 + N \right)} \]

\[d_3(N)= \frac{-32256 - 17408\,N} {{N^4}\,\left( 2 + N \right)}  \]

\[d_4(N)= \frac{ -3086848 - 4038528\,N - 1679424\,{N^2} - 215600\,{N^3}} 
{  {N^5}\,{{\left( 2 + N \right) }^2}\,\left( 4 + N \right) } \]

\[d_5(N)=\frac{  -33383424 - 44140288\,N - 18541952\,{N^2} - 2396160\,{N^3}}
 {  {N^6}\,{{\left( 2 + N \right) }^2}\,\left( 4 + N \right) } \]

\[d_6(N)= \frac{  -4025769984 - 8048769024\,N - 6183804416\,{N^2} 
- 2256738944\,{N^3} -    386061056\,{N^4} - 24592512\,{N^5}}
{  {N^7}\,{{\left( 2 + N \right) }^3}\,\left( 4 + N \right) \,
   \left( 6 + N \right) } \]

\normalsize

For the coefficients which follow it is typographically
more convenient to set $d_r(N)=P_r(N)/Q_r(N)$ and to
 tabulate separately the numerator polynomial $P_r(N)$
 and the denominator polynomial $Q_r(N)$,
\scriptsize

\[P_7(N)=  -38338166784 - 76883050496\,N - 59262616576\,{N^2} 
 - 21695712768\,{N^3} - 
   3720514560\,{N^4} - 237404160\,{N^5}\]

\[ Q_7(N)=  {N^8}\,{{\left( 2 + N \right) }^3}\
   \left( 4 + N \right) \,\left( 6 + N \right) \]

\[P_8(N)=   -22398548705280 - 64579066675200\,N - 78988001214464\,{N^2} - 
   53316161983488\,{N^3} - \]\[ 21632718536704\,{N^4} - 5382539503872\,{N^5} - 
   799715671040\,{N^6} - 64733334688\,{N^7} - 2182679664\,{N^8}\]

\[ Q_8(N)=   {N^9}\,{{\left( 2 + N \right) }^4}\,{{\left( 4 + N \right) }^2}\,
   \left( 6 + N \right) \,\left( 8 + N \right) \]

\[P_9(N)=   -197760671023104 - 570072979439616\,N - 697267355156480\,{N^2} - 
   470715808290816\,{N^3}\]\[ - 191030743859200\,{N^4} 
- 47540867562496\,{N^5} - 
   7064458432512\,{N^6} - 571883674496\,{N^7} - 19283466240\,{N^8}\]

\[ Q_9(N)={N^{10}}\,{{\left( 2 + N \right) }^4}\,{{\left( 4 + N \right) }^2}\,
   \left( 6 + N \right) \,\left( 8 + N \right) \]

\[P_{10}(N)=   -135893648670720000 - 506812733453762560\,N 
 - 837961375615287296\,{N^2}\]\[ - 
   808982617436258304\,{N^3} - 505548566324297728\,{N^4} - 
   214244480842262528\,{N^5} - 62694667322504192\,{N^6}\]\[ - 
   12643889177861632\,{N^7} - 1719173758968320\,{N^8} - 
   149853167848320\,{N^9}\]\[
 - 7525864311296\,{N^{10}} - 164795456384\,{N^{11}}\]

\[ Q_{10}(N)={N^{11}}\,{{\left( 2 + N \right) }^5}\,
{{\left( 4 + N \right) }^3}\,
   \left( 6 + N \right) \,\left( 8 + N \right) \,\left( 10 + N \right) \]

\[P_{11}(N)=   -1141044912712581120 - 4249040663391240192\,N - 
   7015555079356284928\,{N^2} \]\[- 6764393828229578752\,{N^3} - 
   4222404584768831488\,{N^4} - 1787577880615559168\,{N^5}\]\[ - 
   522623903804669952\,{N^6} - 105313027917844480\,{N^7} - 
   14308686372335616\,{N^8}\]\[ - 1246403878745600\,{N^9} - 
   62559888474112\,{N^{10}} - 1369193398272\,{N^{11}}\]

\[Q_{11}(N)={N^{12}}\,{{\left( 2 + N \right) }^5}\,
{{\left( 4 + N \right) }^3}\,
   \left( 6 + N \right) \,\left( 8 + N \right) \,\left( 10 + N \right) \]

\[P_{12}(N)=   -1353224519826548981760 - 6043648174007450075136\,N - 
   12246178519860941684736\,{N^2}\]\[ - 14898334498567895384064\,{N^3} - 
   12137516326724153901056\,{N^4} - 6993670039948027871232\,{N^5}\]\[ - 
   2934699277523969310720\,{N^6} - 909774043355138494464\,{N^7} - 
   209115782694899085312\,{N^8} \]\[- 35427294974940029952\,{N^9} - 
   4350609416485181440\,{N^{10}} - 375257673683951872\,{N^{11}}\]\[ - 
   21477632431608320\,{N^{12}} - 729782591023616\,{N^{13}} - 
   11103491816320\,{N^{14}}\]

\[ Q_{12}(N)=   {N^{13}}\,{{\left( 2 + N \right) }^6}\,
{{\left( 4 + N \right) }^3}\,
   {{\left( 6 + N \right) }^2}\,\left( 8 + N \right) \,
   \left( 10 + N \right) \,\left( 12 + N \right) \]

  \[P_{13}(N)= -10962814317012556185600 - 
   48855640313305403228160\,N - 98790788652478570168320\,{N^2}\]\[ - 
   119949033747113622634496\,{N^3} - 97539319548016125280256\,{N^4} - 
   56104431660229407899648\,{N^5} \]\[- 23504510497602793209856\,{N^6} - 
   7275637584279889346560\,{N^7} - 1670046301220816199680\,{N^8}\]\[ - 
   282577439283542464512\,{N^9} - 34662501972894392320\,{N^{10}} - 
   2986763073027756032\,{N^{11}}\]\[ - 170792230703923200\,{N^{12}} - 
   5798723600550656\,{N^{13}} - 88165648564224\,{N^{14}}\]

\[ Q_{13}(N)=   {N^{14}}\,{{\left( 2 + N \right) }^6}\,
{{\left( 4 + N \right) }^3}\,
   {{\left( 6 + N \right) }^2}\,\left( 8 + N \right) \,
   \left( 10 + N \right) \,\left( 12 + N \right) \]

\[P_{14}(N)= -58832199553811357682892800 - 319656261223016276133150720\,N\]\[
 - 804506884372627522672656384\,{N^2} 
- 1245051701802465742955741184\,{N^3}\]\[
 -  1326506792510663276202295296\,{N^4} - 
   1032281846527637618727845888\,{N^5}\]
\[ - 607550530706546656156057600\,{N^6} - 
   276285749123163236534124544\,{N^7}\]
\[ - 98336895868270792502870016\,{N^8} - 
   27578892092469750017228800\,{N^9}\]
\[ - 6104887252432902432088064\,{N^{10}} - 
   1063450425127326143963136\,{N^{11}}\]
\[ - 144612518832699900119040\,{N^{12}} - 
   15134717223692504156160\,{N^{13}}\]\[ - 1191613756119982431744\,{N^{14}} - 
   68071027536755782912\,{N^{15}}\]\[ - 2655538734547345536\,{N^{16}} - 
   63104921069570176\,{N^{17}} - 687220401024000\,{N^{18}}\]

\[ Q_{14}(N)=   {N^{15}}\,{{\left( 2 + N \right) }^7}\,
{{\left( 4 + N \right) }^4}\,
   {{\left( 6 + N \right) }^3}\,\left( 8 + N \right) \,
   \left( 10 + N \right) \,\left( 12 + N \right) \,\left( 14 + N \right) \]

  \[P_{15}(N)= -463996271577436867233054720 
   - 2515152883396655909638766592\,N - 
   6315643027876160842940547072\,{N^2}\]\[ - 
   9752398559014235653061738496\,{N^3} - 
   10368212405864877478064947200\,{N^4}\]\[ - 
   8051953080310134001281007616\,{N^5} - 
   4729721489677266022278627328\,{N^6}\]\[ - 
   2146872892026044965004247040\,{N^7} 
   - 762790924816930341473615872\,{N^8}\]\[ - 
   213577145756373531731427328\,{N^9} - 
   47205533784466451589431296\,{N^{10}}\]\[ - 
   8211418569550233134678016\,{N^{11}} - 
   1115168266958340559364096\,{N^{12}} \]
\[- 116570698649671407411200\,{N^{13}} - 
   9168049245864669396992\,{N^{14}}\]\[ - 523207897673566349312\,{N^{15}} - 
   20392833008810153984\,{N^{16}} - 484218500284702720\,{N^{17}} - 
   5269451094097920\,{N^{18}}\]

\[ Q_{15}(N)=   {N^{16}}\,{{\left( 2 + N \right) }^7}\,
{{\left( 4 + N \right) }^4}\,
   {{\left( 6 + N \right) }^3}\,\left( 8 + N \right) \,
   \left( 10 + N \right) \,\left( 12 + N \right) \,\left( 14 + N \right) \]

  \[P_{16}(N)= -3706032155932253013818065551360 - 
   23511532957902177899390315790336\,N\]\[ - 
   69984745734719724031602453381120\,{N^2} - 
   129942518358149653545095685734400\,{N^3}\]\[ - 
   168777163612514463301490677121024\,{N^4} - 
   163020601132834974169821018587136\,{N^5}\]\[ - 
   121523005151702351276069798019072\,{N^6} - 
   71619975067826655255089882595328\,{N^7}\]\[ - 
   33913954817349955356957471670272\,{N^8} - 
   13042529099485925377885106077696\,{N^9}\]\[ - 
   4101432698345456358253122813952\,{N^{10}} - 
   1058465798499513773553711382528\,{N^{11}}\]\[ - 
   224369655134353441277932077056\,{N^{12}} - 
   38996967102960612063728975872\,{N^{13}}\]\[ - 
   5532347732173533521822597120\,{N^{14}} - 
   635702582117166696141393920\,{N^{15}}\]\[ - 
   58478148624089490157039616\,{N^{16}} - 
   4233637928944288589514240\,{N^{17}}\]\[
 - 235251240860669046267648\,{N^{18}} - 
   9658642370970465608960\,{N^{19}}\]\[
 - 275363031292904868480\,{N^{20}} - 
   4856360702431679584\,{N^{21}} - 39817256443223280\,{N^{22}}\]

\[ Q_{16}(N)= {N^{17}}\,{{\left( 2 + N \right) }^8}\,
{{\left( 4 + N \right) }^5}\,
   {{\left( 6 + N \right) }^3}\,{{\left( 8 + N \right) }^2}\,
   \left( 10 + N \right) \,\left( 12 + N \right) \,\left( 14 + N \right) \,
   \left( 16 + N \right) \]

  \[P_{17}(N)=   -28625170809317622639273757900800 
   - 181166520457100717483475767132160\,N\]\[ - 
   537991814958099333262399103827968\,{N^2} - 
   996597299051903297408062545985536\,{N^3}\]\[ - 
   1291524895757272450924715899355136\,{N^4} - 
   1244744825531489741899730562056192\,{N^5}\]\[ - 
   925926421634294684791564008423424\,{N^6} - 
   544586262055545477273422842036224\,{N^7}\]\[ - 
   257372711584126360801923714514944\,{N^8} - 
   98795363631395088253841625841664\,{N^9}\]\[ - 
   31012859792452238511372108300288\,{N^{10}} - 
   7990178234700265757218112274432\,{N^{11}}\]\[ - 
   1691062951010009264622059913216\,{N^{12}} - 
   293484614468719284469043789824\,{N^{13}}\]\[ - 
   41578182122239994876120399872\,{N^{14}} - 
   4771486216282518354479939584\,{N^{15}}\]\[ - 
   438407834252226343769522176\,{N^{16}} - 
   31704698511771193169897472\,{N^{17}} \]\[- 
   1759972842290831947573248\,{N^{18}}
 - 72192427708156325499904\,{N^{19}}\]\[ - 
   2056451245211192506368\,{N^{20}} 
- 36240554109317191808\,{N^{21}} - 
   296932880108021760\,{N^{22}}\]

\[ Q_{17}(N)= {N^{18}}\,{{\left( 2 + N \right) }^8}\,
   {{\left( 4 + N \right) }^5}\,{{\left( 6 + N \right) }^3}\,
   {{\left( 8 + N \right) }^2}\,\left( 10 + N \right) \,
   \left( 12 + N \right) \,\left( 14 + N \right) \,\left( 16 + N \right) \]

\normalsize

	In particular for $N=0$, we have (in terms of the variable 
	$\tilde \beta=\beta/N$) 

\scriptsize

\[\hat \chi_4(\tilde\beta)= 
                                -3
                               -96 \tilde\beta 
                              -1728, \tilde\beta^{ 2} 
                            -24192 \tilde\beta^{ 3} 
                           -289392 \tilde\beta^{ 4} 
                          -3129696 \tilde\beta^{ 5} 
                          -31451328 \tilde\beta^{ 6} 
                        -299516928 \tilde\beta^{ 7} 
                       -2734197840 \tilde\beta^{ 8} \]\[
                      -24140706912 \tilde\beta^{ 9} 
                     -207357252000 \tilde\beta^{10} 
                    -1741096363392 \tilde\beta^{11} 
                    -14339282987664 \tilde\beta^{12} 
                    -116166160551840 \tilde\beta^{13} \]\[
                   -927691099407360\tilde\beta^{14} 
                 -7316490196952064 \tilde\beta^{15} 
                -57068641022992368 \tilde\beta^{16} 
               -440794771444139040 \tilde\beta^{17}\]

\normalsize

	For $N=1$ [ the spin 1/2 Ising model], we have 

\scriptsize

\[ \chi_4(1,\beta)=
-2 -64\beta  -1168\beta^{2 } -49664/3\beta^{ 3} -601360/3\beta^{ 4}
 -32820608/15\beta^{ 5} -996463616/45\beta^{ 6} 
  -66712488448/315\beta^{ 7}\]\[ -122056132496/63\beta^{ 8} 
 -48489867797888/2835\beta^{ 9} 
  -2078558044733696/14175\beta^{10 }\]\[  
  -191285725186144768/155925\beta^{ 11} 
  -188087379936809600/18711\beta^{12 }
  -492034524872707515136/6081075\beta^{13 } \]\[
  -27296494302637993572352/42567525\beta^{ 14} 
 -3201197677867739316248576/638512875\beta^{ 15} \]\[
  -4945781553886665074906384/127702575\beta^{ 16} 
  -3212941768987291424807915648/10854718875\beta^{ 17}\]

\normalsize

	For $N=2$ [ the XY model], we have 

\scriptsize
\[ \chi_4(2,\beta)= -3/2 -24\beta -441/2\beta^{2 } -1572\beta^{3 }
 -153175/16\beta^{4 }
 -104981/2\beta^{5 } -68238647/256\beta^{6 } \]\[
  -40884131/32\beta^{7 }  -1404217891/240\beta^{8 }
 -66125311269/2560\beta^{9 }
 -16313466298147/147456\beta^{10 }\]\[ 
  -85154694896333/184320\beta^{11 }
 -51968444431571323/27525120\beta^{12 } 
  -156353752523792639/20643840\beta^{13 } \]\[
-236982809408746803649/7927234560\beta^{14 }
  -400259785750849937/3440640\beta^{15 }\]\[
 -637876955227851294690449/1426902220800\beta^{16 }
  -48476858619011264835409/28538044416\beta^{17 }...\]

\normalsize

	For $N=3$ [the Heisenberg classical model], we have 
\scriptsize

\[ \chi_4(3,\beta)=
-6/5 -64/5\beta  -1968/25\beta^{2 } -5632/15\beta^{ 3}
 -36138448/23625\beta^{ 4} -132459392/23625\beta^{ 5}
  -1348186624/70875\beta^{ 6} \]\[ -194131778048/3189375\beta^{ 7}
 -228357648983536/1227909375\beta^{ 8} 
  -2016513048715136/3683728125\beta^{ 9} \]\[
-3366834959446902016/2154980953125\beta^{ 10} 
  -1872724144398680576/430996190625\beta^{ 11}\]\[
 -2674610910995288182912/226273000078125\beta^{ 12} 
  -367131807235133274368/11636897146875\beta^{13 } \]\[
  -100350171877191930199645952/1211691915418359375\beta^{ 14} 
  -778819344453674012213276672/3635075746255078125\beta^{ 15} \]\[
  -130937251648404408278387043967856/239315211754703068359375\beta^{16} \]\[
  -22035313880683966424477522278528/15954347450313537890625\beta^{17 }\]


\begin{references}

\bibitem[*]{pb}Electronic address: butera@mi.infn.it
\bibitem[**]{mc}Electronic address: comi@mi.infn.it

\bibitem{st68} H. E. Stanley,  Phys. Rev. Lett. {\bf20}, 589 (1968);
 and in {\it Phase Transitions and Critical Phenomena}, edited by C. Domb
and M. S. Green,(Academic, New York,1974) Vol.3.


\bibitem{bc2d}P.Butera and M. Comi, Phys. Rev. B {\bf 54}, 15828 (1996).
\bibitem{bc3d}P.Butera and M. Comi, Phys. Rev. B {\bf 56}, 8212 (1997).

 \bibitem{gut87} A. J. Guttmann, J. Phys. A {\bf 20},  1855 (1987).

\bibitem{lw88} M. L\" uscher and P.Weisz
Nucl. Phys. B {\bf 300}, 325  (1988). 

\bibitem{b90} P. Butera, M. Comi, and G. Marchesini,
Phys.Rev. B {\bf 41}, 11494 (1990);
 P. Butera,  M. Comi and G. Marchesini,
Nucl. Phys. B {\bf 300}, 1 (1988).

\bibitem{w74mck83} F. Englert,  Phys.Rev. {\bf 129}, 567 (1963);
M. Wortis, D. Jasnow and M. A. Moore, Phys.Rev. {\bf 185}, 805 (1969); 
M. Wortis, in {\it Phase Transitions and critical Phenomena}, 
edited by C. Domb and M. S. Green, (Academic,
London, 1974), Vol. 3;  S. McKenzie,
  in {\it Phase Transitions: Cargese 1980}, edited by
M. Levy, J. C. Le Guillou and J. Zinn Justin, (Plenum, New York,
1982).
\bibitem{bk}
G. A. Baker, H. E. Gilbert, J. Eve and G. S. Rushbrooke {\it  A data 
compendium of linear graphs with application to the Heisenberg model}, 
Brookhaven National Laboratory  unpublished 
report BNL 50053 (T-460) (1967);  J. M. Kincaid, G. A. Baker
 and L. W. Fullerton {\it High temperature series expansions of the 
continuous spin Ising Model} ;  Los Alamos unpublished report 
LA-UR-79-1575 (1979).

\bibitem{bakeb} G. A. Baker, {\it Quantitative
Theory of Critical Phenomena}, (Academic, Boston, 1990).

\bibitem{rs} R. Z. Roskies and P. D. Sackett,
J. Stat. Phys. {\bf 49}, 447 (1987).

\bibitem{re} A computation of $\chi_4$ to order 
$\beta^{16}$ only on the sc lattice
has been announced in Ref.\cite{r1}, but no series have yet 
been published.

\bibitem{r1} T. Reisz, Phys. Lett. B {\bf360} (1995) and
 Nucl. Phys. Proc. Suppl. {\bf 53}, 841 (1997).

\bibitem{mackenzie} S. Mackenzie, Can. J. Phys.,
{\bf 57}, 1239 (1979).


\bibitem{gutt} A. J. Guttmann, Phys. Rev. B {\bf33} 5089 (1986). 


\bibitem{ns} B.G.Nickel and B. Sharpe, J. Phys. 
A {\bf 12}, 1818 (1979).

\bibitem{roskies}  R. Roskies, Phys. Rev.  B {\bf 23}, 6037 (1981).



\bibitem{essam} J.W. Essam and D. L. Hunter, J. Phys. 
C {\bf1}, 392 (1968); M. A. Moore, D. Jasnow and M. Wortis,
 Phys. Rev. Lett. {\bf 22}, 940 (1969).

\bibitem{stan2}H.E. Stanley  Phys. Rev. {\bf 176}, 718 (1968).

\bibitem{joy}  G. S. Joyce, in  {\it Phase Transitions and
Critical Phenomena}, edited by C. Domb
and M. S. Green,(Academic, New York,1972) Vol.2; 
 Phil. Trans. Roy. Soc. {\bf 273}, 583 (1973).


\bibitem{mur} D. B. Murray and B. G. Nickel, {\it Revised estimates 
for critical exponents  for the continuum N-vector model in 3 dimensions},
Unpublished Guelph University report (1991).

\bibitem{gz2} R. Guida and J. Zinn-Justin,{\it Critical exponents of 
the $N-$vector model}, CEA-Saclay Preprint SPhT-t97/040, cond-mat/9803240.  

\bibitem{zib} G.Parisi, Cargese 1973 Lecture Notes, unpublished;
 J. Stat. Phys. {\bf 23}, 49 (1980); 
{\it Statistical Field Theory} (Addison Wesley, New York, 1988).

\bibitem{zib1}
G.A. Baker, B. G. Nickel M. S. Green and D. I. Meiron, Phys. Rev. Lett.
 {\bf 36}, 1351 (1976);
G.A. Baker, B. G. Nickel and D. I. Meiron, Phys. Rev. B {\bf 17}, 
1365 (1978);
J. C. Le Guillou and J. Zinn Justin, 
Phys. Rev. Lett. {\bf 39}, 95 (1977).


  
\bibitem{brez} E. Brezin, J. C. Le Guillou 
and J. Zinn-Justin, in  {\it Phase 
Transitions and critical Phenomena}, edited by
C. Domb and M. S. Green, (Academic, New York, 1976) Vol. 6.


\bibitem{zinn}  J. Zinn Justin, {\it Quantum field theory and critical
phenomena} (Clarendon, Oxford, 1989, third edition 1996).

\bibitem{itdr} C. Itzykson and J. M. Drouffe, {\it Statistical Field Theory},
(Cambridge University Press, 1989).


\bibitem{ant}S. A. Antonenko and A. I. Sokolov,  Phys. Rev. E {\bf 51}, 
1894 (1995). 


\bibitem{rogiers} D.M. Saul, M. Wortis and D. Jasnow, Phys. Rev. B {\bf 11},
 2571 (1975); 
 W. J. Camp,  D.M. Saul, J. P. Van Dyke and M. Wortis, 
Phys. Rev. B {\bf 14}, 3990 (1976);
 W. J. Camp and J. P. Van Dyke  Phys. Rev. B {\bf 19}, 2579 (1979); 
J. Rogiers, M. Ferer and E. R. Scaggs, 
Phys. Rev. B {\bf 19}, 1644 (1979).


\bibitem{gaunt80} D. S. Gaunt, in 
{\it Phase Transitions: Cargese 1980}, edited by M. Levy , J. C. Le Guillou
and J. Zinn-Justin, (Plenum,  New York, 1982).

\bibitem{camp80} W. J. Camp, in 
{\it Phase Transitions: Cargese 1980}, edited by M. Levy , J. C. Le Guillou
and J. Zinn-Justin, (Plenum,  New York, 1982).


\bibitem{nickel80} B.G.Nickel,
in {\it Phase Transitions: Cargese 1980}, 
edited by M. Levy , J. C. Le Guillou
and J. Zinn-Justin, (Plenum,  New York, 1982).


\bibitem{zinn79}J. Zinn  Justin, J. Physique {\bf40}, 969 (1979); 
 ibid. {\bf42}, 783 (1981)

\bibitem{adl} J. Adler, J. Phys. A  {\bf 16}, 3585 (1983). 

\bibitem{fv} M. Ferer and M. J. Velgakis, Phys. Rev. B  {\bf 27}, 2839 (1983). 

\bibitem{fisher}J.H. Chen, M. E. Fisher and B. G. Nickel, 
 Phys. Rev. Lett. {\bf 48}, 630 (1982);  M. E. Fisher and J.H. Chen, 
J. Physique  {\bf 46}, 1645 (1985).

\bibitem{nr90} B.G.Nickel and J.J. Rehr,
J. Stat. Phys. {\bf 61}, 1 (1990).


\bibitem{liu1} A. E. Liu and M. E. Fisher, Physica A
{\bf156}, 35 (1989).


\bibitem{liufi} A. E. Liu and M. E. Fisher, J. Stat. Phys. 
{\bf58}, 431 (1990).


\bibitem{wegner} F. Wegner, Phys. Rev. B  {\bf 5}, 4529 (1972).

\bibitem{aha}V. Privman, P. C. Hohenberg and A. Aharony, 
 in {\it Phase Transitions and critical Phenomena}, edited by 
C. Domb and J. Lebowitz, (Academic, New York, 1989) Vol. 14;
P. C. Hohenberg,  A. Aharony, B. I. Halperin and E.D. Siggia 
Phys. Rev. B  {\bf 13}, 2896 (1976);  A. Aharony and P. C. Hohenberg,
Phys. Rev. B  {\bf 13}, 3081 (1976).

\bibitem{epsi}
 M. E. Fisher and K.G. Wilson, Phys. Rev. Lett. {\bf28}, 240 (1972).
\bibitem{epsi1}
A. A. Vladimirov, D. I. Kazakov and O.V. Tarasov, 
 Sov. Phys. JEPT {\bf50}, 521 (1979), K. G. Chetyrkin, 
A. L. Kataev and F. V. Tkachov, Phys. Lett. B{\bf99}, 147 (1981); 
 B {\bf101}, 457(E) (1981);
K. G. Chetyrkin and F. V. Tkachov, Nucl. Phys.  B {\bf192}, 159 (1981);
K. G. Chetyrkin, S. G. Gorishny, S. A. Larin and F. V. 
Tkachov, Phys. Lett. B {\bf132}, 351 (1983);
D. I. Kazakov, Phys. Lett. B {\bf133}, 406 (1983); 
S. G. Gorishny, S. A. Larin and F. V. 
Tkachov, Phys. Lett. A {\bf101}, 120 (1984); H. Kleinert, J. Neu, V. 
Schulte-Frohlinde and S. A. Larin, Phys. Lett. B {\bf272}, 39 (1991); 
B {\bf319}, 545(E) (1993). 
 
\bibitem{epv} J. C. Le Guillou and J. Zinn Justin, 
Phys. Rev. B {\bf 21}, 3976 (1980);
J. C. Le Guillou and J. Zinn Justin, J. Physique Lett. {\bf46}, 137 (1985);
J. Physique {\bf50}, 1365 (1989); J. Zinn Justin,
in {\it Phase Transitions: Cargese 1980}, 
edited by M. Levy , J. C. Le Guillou
and J. Zinn-Justin, (Plenum,  New York, 1982).

\bibitem{pv} A. Pelissetto and E. Vicari, IFUP-TH 52/97, cond-mat/9711078.

\bibitem{bgz} E. Brezin, J. C. Le Guillou and J. Zinn-Justin, 
Phys. Rev. D {\bf 8},  434 (1973).


\bibitem{gen} P.G. de Gennes, Phys. Lett. A {\bf38}, 339 (1972).


\bibitem{ma} S. K. Ma,  Phys. Rev. A  {\bf 10}, 1818 (1974). 

\bibitem{tarko} M. E. Fisher and R. J. Burford,  Phys. Rev.
  {\bf 156}, 583 (1966).

\bibitem{fernandez} A. Fernandez, J. Froehlich and A. Sokal, {\it
Random walks, critical phenomena and triviality in quantum field theory},
Springer Verlag, Berlin 1992.

\bibitem{lms} B. Li, N. Madras and A. D. Sokal, J. Stat. Phys. 
{\bf 80}, 661 (1995).


\bibitem{gunt} J.D. Gunton and M.J. Buckingam, Phys. 
Rev. {\bf 178}, 848 (1969).

\bibitem{rig}  Rigorous studies of this inequality appear in Ref.\cite{glja}.

\bibitem{glja} J. Glimm and A. Jaffe, Ann. Inst. H. Poincare' 
A{\bf22}, 97 (1975); 
 R. Schrader,  Phys. Rev. B {\bf 14}, 172 (1976); 
 M. Aizenman, Comm. Math. Phys. {\bf 86}, 1 (1982); 
E.H. Lieb and A. D. Sokal, 
unpublished results quoted in Ref.\cite{fernandez}. 

\bibitem{lebo} J. Lebowitz, Comm. Math. Phys. {\bf 35}, 87 (1974).

\bibitem{fereconf} M. Ferer, Phys. Rev. B {\bf 16}, 419 (1977).

\bibitem{gutasy}
A. J. Guttmann and G. S. Joyce, J. Phys. A 
{\bf 5}, L81 (1972);
D. L. Hunter and G. A. Baker, Phy. Rev.
 B {\bf 19}, 3808 (1979); M. E. Fisher and H. Au-Yang, J. Phys. A 
{\bf 12}, 1677 (1979) and A {\bf 13}, 1517 (1980); 
 J. J. Rehr, A. J. Guttmann and G. S. Joyce, ibid.
{\bf 13}, 1587 (1980);  
 A. J. Guttmann,
 in {\it Phase Transitions and critical Phenomena}, edited by 
C. Domb and J. Lebowitz, (Academic, New York, 1989) Vol. 13.

\bibitem{bcpre}P.Butera and M. Comi, Phys. Rev. E {\bf 55}, 6391 (1997).


\bibitem{sokoth} A. I. Sokolov, to be published in
 Fizika Tverdogo Tela, {\bf40} (1998).


\bibitem{rosk} R. Roskies, Phys. Rev.  B {\bf 24}, 5305 (1981).

\bibitem{adp}  J. Adler, M. Moshe and V. Privman,
 Phys. Rev. B{\bf 26}, 3958 (1982).
\bibitem{nd}B. G. Nickel and M. Dixon, Phys. Rev.  B {\bf 26}, 3965 (1982).

\bibitem{nick81} B. G. Nickel, Physica A {\bf 117}, 189  (1981).



\bibitem{bahu}G. A. Baker and D. L. Hunter, 
Phys. Rev. B {\bf 7},  3377 (1973).

\bibitem{def} P. de Forcrand, F. Koukiou and D. Petritis,  
Phys. Lett. B{\bf189} , 341 (1987);  J. Stat. Phys. {\bf 49}, 223 (1987).

\bibitem{dombook} C. Domb, {\it The critical point}, 
(Taylor and Francis, London 1996).


\bibitem{bake} G. A. Baker, Phys.Rev. B {\bf 15}, 1552 (1977);
  G. A. Baker and J. M. Kincaid,
J. Stat. Phys. {\bf 24}, 469 (1981).





\bibitem{rehr} J. J. Rehr, J. Phys. A  {\bf 12}, L179 (1979).  


\bibitem{tsy} M. M. Tsypin, Phys. Rev. Lett. {\bf 73}, 2015 (1994). 

\bibitem{kl} J. K. Kim and A. Patrascioiu Phys. Rev. D {\bf 47},  2588 (1993).
J. K. Kim and D.P.  Landau, Nucl. Phys. Proc. Suppl. {\bf 53}, 706 (1997).



\bibitem{baka} G.A. Baker Jr. and N. Kawashima, 
Phys. Rev. Lett. {\bf 75}, 994 (1995); J. Phys. A {\bf29}, 7183(1996).

\bibitem{causo}S. Caracciolo, M. S.  Causo, A. Pelissetto, cond-mat/9703250,
 Nucl. Phys. B (Proc. Suppl.) {\bf 63} A-C, 652 (1998). 


\bibitem{zlf} S. Y. Zinn, S. N. Lai and  M. E. Fisher,
Phys. Rev. E {\bf54}, 1176 (1996).

\bibitem{rzw}C.Ruge, P.Zhou and F. Wagner, Physica A {\bf209}, 431(1994).

\bibitem{got} A. P. Gottlob and M. Hasenbusch, Physica A {\bf201}, 593 (1993).
 
\bibitem{che} C. Holm and W. Janke, Phys. Rev. B {\bf 48}, 936 (1993).


\bibitem{george} M. J. George and J.J. Rehr, Phys. Rev. Lett. {\bf53}, 
2063 (1984); M. J. George, Washington University 1985 Ph.D. thesis.



\bibitem{ch} M. C. Chang and M. C. Houghton , 
Phys. Rev. Lett. {\bf44}, 785 (1979).

\bibitem{cr} M. C. Chang and J. J. Rehr, J. Phys. A {\bf16}, 3899 (1983).


\bibitem{bb} C. Bagnuls and C. Bervillier,  
Phys. Rev. B {\bf 24}, 1226 (1981); 
J. Phys. A {\bf19}, L85 (1986).
 
\bibitem{blh} H. W. J. Bl\" ote, E. Luijten and J. R. Heringa, J. Phys. 
A {\bf 28}, 6289 (1995); A. L. Talapov and H. W. J. Bl\" ote, J. Phys. 
A {\bf 29}, 5727 (1996).


\end{references}
\end{document}